\documentclass{aa}  
\usepackage{graphicx,subcaption}
\usepackage{txfonts}
\usepackage[]{hyperref}
\usepackage{enumitem}
\usepackage{lscape}
\usepackage[T1]{fontenc}
\newcommand*{\ditto}{--- \raisebox{-0.5ex}{''} ---}

\begin{document} 

   \title{Finding the elusive RR Lyrae companions via speckle imaging}

   \author{R. Salinas \inst{1} \and  V. Kalari  \inst{2} \and G. Hajdu \inst{1} \and Z. Prudil \inst{3} \and C. Sáez-Carvajal\inst{4} \and W. Narloch\inst{1} \and M. Catelan\inst{5,6} \and S.B. Howell\inst{7} \and K. Bąkowska\inst{8} \and R. Chini\inst{1,9} \and C. Gałan\inst{1} \and M. Górski\inst{1} \and M. Kałuszyński\inst{1} \and P. Karczmarek\inst{1} \and M. Kicia\inst{1} \and W. Kiviaho\inst{10,11} \and K. Kotysz\inst{12,13} \and F. Marcadon\inst{1} \and D. Moździerski\inst{13} \and H. Netzel\inst{1} \and G. Pietrzyński\inst{1} \and W. Pych\inst{1} \and M. Radziwonowicz\inst{1} \and P. Romaniuk\inst{1} \and R. Smolec\inst{1} \and P. Wielgórski\inst{1} \and B. Zgirski\inst{14} \and P. Żuk\inst{1}}

   \institute{Nicolaus Copernicus Astronomical Center, Polish Academy of Sciences, Bartycka 18, 00-716 Warszawa, Poland \\ \email{rsalinas@camk.edu.pl} \and Gemini Observatory/NSF's NOIRLab, Casilla 603, La Serena, Chile \and European Southern Observatory, Karl-Schwarzschild-Strasse 2, 85748, Garching bei München, Germany \and Instituto de Física y Astronomía, Universidad de Valparaíso, Av. Gran Bretaña 1111, 5030 Casilla, Valparaíso, Chile \and Institute of Astrophysics, Pontificia Universidad Católica de Chile, Av. Vicuña Mackenna 4860, 7820436, Macul, Santiago, Chile \and Centro de Astroingeniería, Pontificia Universidad Católica de Chile, Av. Vicu\~na Mackenna 4860, 7820436 Macul, Santiago, Chile \and NASA Ames Research Center, Moffett Field, CA 94035, USA \and Institute of Astronomy, Faculty of Physics, Astronomy and Informatics, Nicolaus Copernicus University in Toruń, Grudziądzka 5, 87-100 Toruń, Poland \and Ruhr University Bochum, Faculty of Physics and Astronomy, Astronomical Institute (AIRUB), 44780 Bochum, Germany \and LIRA, Observatoire de Paris, Université PSL, Sorbonne Université, Université Paris Cité, CY Cergy Paris Université, CNRS, 5 place Jules Janssen, 92195 Meudon, France \and French-Chilean Laboratory for Astronomy, IRL 3386, CNRS and U. de Chile, Casilla 36-D, Santiago, Chile \and Astronomical Observatory, University of Warsaw, Aleje Ujazdowskie 4, 00-478 Warsaw, Poland \and Astronomical Institute, University of Wrocław, ul. Mikołaja Kopernika 11, 51-622 Wrocław, Poland \and Universidad de Concepción, Departamento de Astronomía, Casilla 160-C, Concepción, Chile}

   \date{}

  \abstract
   {}
   {Despite their key role in astrophysics, the binary properties of RR Lyrae stars (RRL) remain almost completely unknown since only a single RRL is confirmed as belonging to a binary system. Finding companions to RRL is difficult since most of them will be at wider orbits,  given that close orbits will likely ensue mass transfer disrupting the conditions to develop stellar pulsations. These wide orbits open the possibility that RRL companions may be more easily found by high-resolution imaging.}
   {We observed 81 RRL with the speckle interferometers Zorro and 'Alopeke at the Gemini telescopes, reaching the diffraction limit of $\sim$20 mas of these 8m-class telescopes, and therefore exploring a new parameter space around RRL.}
   {We have detected 10 newly identified companions around these 81 RRL, with projected separations between 20 AU to 220 AU. An analysis of the field contamination shows that all of these  detected companions are most likely gravitationally bound binaries. From these observations we can estimate an RRL binary fraction higher than 12\%, ruling out a binary fraction higher than 25\% at the 99\% confidence level. These numbers are significantly more elevated than previous estimations which were close to a binary fraction of only 1\%,  albeit derived with methods exploring a different parameter space. For RRL with thin disc kinematics,  we find that the binary fraction is significantly lower, at around 6\%, with a single thin disc RRL having a companion  out of the 16 observed. The nature of the companions,  found to be stars in the lower red giant branch and upper main sequence, is also studied via the measurement of the minimum light colors of the RRL, which appears as a useful method for the search and analysis of RRL in binary systems.}   
   {}


\maketitle 
\nolinenumbers
\section{Introduction}\label{sec:intro}

Arguably, the most important physical quantity of any star is its mass \citep[e.g.][]{eddington24,vogt26}, which determines their luminosity, temperature, inner structure, and lifetime. Unsurprisingly then, measuring stellar masses is, to this day, one of the core endeavours of observational astronomy. However, there are very few methods that allow for a direct measurement of stellar masses, the most accurate being the study of eclipsing or astrometric binary stars, where the joint analysis of photometric and/or astrometric together with spectroscopic observations can provide masses with a precision better than 1\% \citep[e.g.][]{southworth12}.

One advantage is that binaries are plentiful across the Hertzsprung-Russell diagram \citep[e.g.][]{duchene13}. The fraction of binary systems among early-type massive stars is as high as 70\% \citep{sana12}, and although this fraction decreases towards low-mass stars, for M dwarfs the multiplicity is still about 20\% \citep{winters21}.

Despite this widespread existence of binaries, the binary fraction of one stellar population has remained particularly elusive: RR Lyrae stars (hereafter RRL). RRL are horizontal branch pulsating stars, with periods between 0.2 and 0.9 days, which provide important constraints for aspects of stellar structure and pulsation theory \citep{catelan15}. They also follow period-luminosity relations making them one of the prime distance indicators for the oldest stellar populations \citep{beaton18}.

 Yet the lack of RRL in binaries means their masses have been inferred mostly indirectly via double-mode RRL \citep[e.g.][]{bono96,netzel22}, but not directly through orbit modelling. To date, only a \textit{single} RRL has been confirmed as member of a binary system \citep[TU UMa,][]{wade99,liska16tu}, but if we consider that the main sequence progenitors of RRL are late G stars or early K stars, then naively we would expect up to 40\% of RRL to reside in binary systems \citep[e.g.][]{raghavan10}.

The striking difference between the expected amount of RRL in binary systems and the almost null detection of such systems has prompted several searches of RRL in binary systems. One of the most traditional ways to find binaries is via radial velocity (RV) searches, which in the case of RRL requires disentangling the RV signal from the pulsation from the orbital motion, requiring observations during many epochs \citep[e.g][]{guggenberger16}. Even though some binary candidates have been proposed in the past via this method \citep[e.g.][]{fernley97,solano97}, more modern dedicated searches have come back empty handed \citep{barnes21,poretti25}, although they can provide important upper limit constraints for the companion masses.

A second search method is the study of the proper motion anomaly (herefater PMa), that is, looking for variations in the proper motion of a star between two distant epochs. These variations or anomalies can be attributed to the existence of a close companion. The method has been applied by \citet{kervella19a} comparing \textsc{Hipparcos} \citep{perryman97} and \textit{Gaia} Data Release 2 \citep{gaiadr2} proper motions to a sample of near 200 RRL, finding significant anomalies, and therefore binary candidates, in 13 cases. Recently, more RRL presenting a PMa have been identified by \citet{abdollahi25}.

The most successful method so far has been the study of $O-C$ diagrams interpreted as a light-time travel effect \citep[LTTE,][]{irwin52}, where the existence of a companion would imprint a periodic wobble in the timing of the RRL light curve, associated with the orbital period of the system. About a hundred RRL in candidate binary systems have been found this way \citep[e.g.][]{liska16lite,prudil19,hajdu21}, albeit  both long period changes \citep[e.g.][]{skarka18,li22} and the Bla\v{z}ko effect \citep[also know as Tseraskaya-Bla\v{z}ko,][]{blazhko07},  present in about 70\% of RRL \citep{molnar22,kovacs25}, may masquerade as LTTE. A number of RR Lyrae in binary system candidates proposed in older studies, mostly through $O-C$ analyses, are discussed in \citet{liska16lite} and compiled in the RRLyrBinCan database\footnote{\href{https://rrlyrbincan.physics.muni.cz/}{https://rrlyrbincan.physics.muni.cz/} (last updated in 2019)} \citep{rrlyrbincan}.

Why finding bonafide RRL in binary systems has been so challenging? The answer is not clear, but it is likely twofold. On the one hand, close binaries would develop mass transfer when the primary or secondary reaches the red giant branch (RGB) phase, a phenomenon that would break the conditions required for RRL pulsation. On the other hand, it is possible that companions of RRL are destroyed when the RRL was in the RGB phase. In both cases companions to RRL will be expected at wider orbits \citep{karczmarek17}. 

One still unexploited method, particularly suited for the search of such wide binaries, is speckle interferometry, where very short exposures close to the atmospheric coherence time aim at ``freezing'' the atmospheric turbulence, allowing the recovery of the diffraction limit of the telescope after processing in the Fourier space \citep{labeyrie70}. Speckle observations can fill the gap between companions detected by RV searches and those binaries directly visible by \textit{Gaia} and  seeing-limited observations \citep[e.g.][]{kalari25}.

 A priori, there are no constraints on the nature of the RRL companions, and their progenitors could in principle have any initial mass. While companions originally more massive than the RRL progenitor would now be compact remnants (white dwarfs, neutron stars or black holes), those with similar and lower masses can be in any other evolutionary stage from the main sequence to the horizontal branch itself. While compact remnants will be not visible by speckle interferometry, the latter should be visible down to 5 to 7 mag below the horizontal branch; the typical contrast achieved with speckle interferometry \citep[e.g.][]{scott21}. One further advantage of speckle interferometry, especially when observations are carried out near minimum light, is that it will remain mostly unaffected by the Bla\v{z}ko effect which complicates the interpretation of $O-C$ diagrams.

\citet{salinas20} observed RRL UV Oct using this method searching for companions between 10 and 633 AU from the star. This paper expands on the work of \citet{salinas20}, with speckle observations of 80 further RRL. Sect. \ref{sec:obs} describes those observations, while Sect. \ref{sec:results} presents the results, and a discussion is given in Sect. \ref{sec:discussion}. A summary and conclusions are provided in Sect. \ref{sec:conclusions}.

\section{Observations and data reduction} \label{sec:obs}

\subsection{The sample} \label{sec:sample}

Given the lack of well-characterized complete volume-limited samples of RRL, we opted instead to focus on the solar neighbourhood sample of \citet{prudil20}. This sample contains 314 fundamental mode RRL with complete information on their metallicities, light curves, radial velocities, proper motions and computed orbits. From this sample we  selected  a subsample of 115 RRL, restricting the original sample to those with mean magnitudes brighter than $V$=11.5 mag and with minimum airmass lower than 1.4 from Maunakea and Cerro Pach\'on. Of these 115 targets, eventually 67 were observed with speckle imaging (see Sect. \ref{sec:speckle}). 

To this original sample, we added a number of RRL which already present some evidence of binarity from the literature, including UV Oct already observed by \citet{salinas20}, here presented in a more detailed analysis and with additional observations, TU UMa, considered as the only confirmed RRL in a binary system \citep[e.g.][]{liska16tu}, and BB Vir, with a suspected blue horizontal branch companion \citep{kinman92}. Additionally, speckle observations were made of 11 further RRL, identified as presenting a significant PMa by \citet{kervella19a} and those from \citet{abdollahi25} who identified RRL from \citet{kervella19a} with thin disc kinematics from \citet{iorio21}. In total, we present speckle observations of 81 RRL.

Main properties of our program RRL, as well as details about their observations can be seen in Table \ref{tab:sample}.

\subsection{Speckle observations} \label{sec:speckle}

Observations were conducted with the speckle interferometers, 'Alopeke and Zorro \citep{scott21}, mounted at the 8.1m Gemini Telescopes in Maunakea, Hawai'i and Cerro Pach\'on, Chile, reaching the diffraction limit of the telescopes of $\sim$20 mas at 500 nm. Zorro and 'Alopeke are equipped with dual Andor EMCCD cameras, with negligible read-out noise and read-out time, allowing for simultaneous observations in two filters. Observations were obtained with medium band filters centered at 562 nm and 832 nm (hereafter EO562 and EO832), with the exception of those targets observed in July 2021 with Zorro, when the blue camera was out of service and only 832 nm data were taken.  Medium band filters are used to minimize the impact of differential atmospheric refraction which can be substantial at these sub-arcsecond scales. Observations were carried out under Gemini programs GS-2019A-SV-401, GN-2020B-FT-115, GN-2020B-LP-105, GS-2021A-Q-220, GN-2021B-Q-309, GS-2021B-Q-315, GN-2025A-FT-211 and GS-2025A-FT-107.

The standard mode of operation of Zorro and 'Alopeke is to acquire individual exposures of 60 ms, taken in groups of 1\,000 exposures. Each of these 1\,000 exposures is called a set. For the majority of the targets, 8 sets were obtained, expected to provide a contrast of $\sim$5 magnitudes at 0.1\arcsec\,from the targets for the typical seeing conditions at Cerro Pachón and Maunakea \citep{howell22}. The exact number of sets taken for each target can be seen in Table \ref{tab:sample}. Only a 256$\times$256 pixel detector window is read, yielding 2.5\arcsec on a side, corresponding to the isoplanatic patch where speckle observations are well defined. All science observations were immediately followed by the observation of a nearby PSF standard that is used for data reduction.

Particular care was taken to observe RRL  on the faint end of the descending branches of their light curves to increase the contrast of the speckle observations.  The timing windows for the observations were selected visually on a case by case basis using the almost concurrent folded light curves from either the ASAS-SN \citep{kochanek17} and/or ATLAS \citep{tonry18} surveys. Timing windows ranged from 25 minutes up to 2.5 hours, and were also constrained by the need to conduct observations at airmasses lower than $\sim$1.4 to minimize the effect of differential refraction given the lack of atmospheric dispersion correctors in the instruments. In a few cases, observations were repeated either because the seeing was considered inadequate, the timing window of the observation was slightly missed, or just to repeat observations that were taken with a single camera. Table \ref{tab:sample} lists both instances.

Speckle data are pipeline reduced following the prescriptions given by \citet{howell11}. Briefly, the power spectrum of each target is calculated by taking the Fourier transform of the summed autocorrelation, dividing by the power spectrum of the PSF standard. If a companion is detected, the fringes are fitted to calculate the separation, magnitude difference, and position angle of the companion relative to the primary star.  These quantities are the main products from speckle observations, and the individual detections are discussed in Sect. \ref{sec:companions}. The pipeline also produces reconstructed images of the  system via bi-spectrum analysis \citep{weigelt77}, and 5-$\sigma$ contrast curves that correspond to the faintest companions detectable at a given separation.  Contrast curves and reconstructed images are produced for all observations, independently on whether a companion is  detected or not. An example of these contrast curves is given in Fig. \ref{fig:contrast}, while those for the rest of the sample are provided in Appendix \ref{app:contrast}.  Reconstructed  images for the detected companions to be discussed in Sect. \ref{sec:companions} are shown in Fig. \ref{fig:reconstructed}.

The photometric accuracy of speckle observations depends mostly on whether the target star and its companion share the same isoplanatic patch. \citet{horch01} introduced the quantity $q'$=seeing$\times\rho$, with $\rho$ being the separation between the two stars in arcsec, as a metric of the degree of correlation of the speckle patterns of the two stars. This metric, as well as the photometric accuracy, were validated with Hipparcos photometry of close binaries. \citet{horch11} found that the photometric accuracy is of the order of 0.1 mag as long as $q'<0.6$ arcsec$^2$, and we use this photometric uncertainty for the detected companions (see Sect. \ref{sec:companions}). Similar values were obtained when comparing with \textit{Gaia} instead of Hipparcos \citep{horch21}. For companions approaching the diffraction limit ($\rho\lesssim50$ mas) the stars in comparison with Hipparcos are scarce, but the photometry is expected to be worse as gleaned from \citet{horch11} (their Fig. 5). For the stars detected at such separations we conservatively assume a photometric uncertainty of 0.3 mag. 

\begin{figure}
\centering
\includegraphics[scale=0.68]{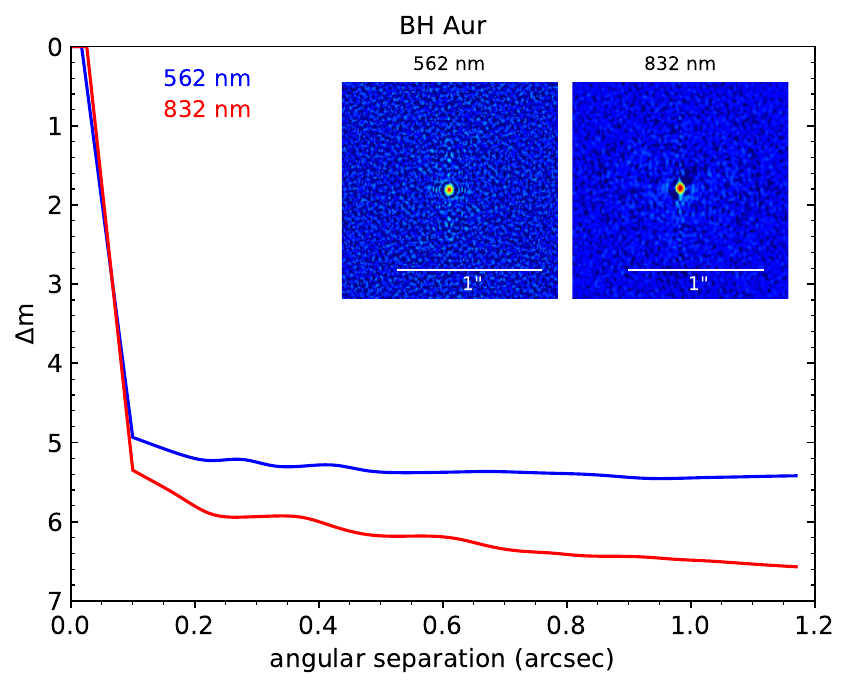}
\caption{5-$\sigma$ contrast curves from speckle observations of RRL BH Aur in both filters EO562 and EO832, together with reconstructed images shown as insets. The shape of the contrast curves is the typical for speckle observations, with a sharp decline from the diffraction limit until 0.1\arcsec, followed by a more gentle decline until the end of the fov. Contrast curves in the redder bandpasses are always deeper given the relatively lower impact of turbulence. Observations of BH Aur reveal no companions above the contrast curves.  Contrast curves for the rest of the observed targets are given in Appendix \ref{app:contrast}.}
    \label{fig:contrast}
\end{figure}

\begin{figure*}
\centering
\includegraphics[scale=0.68]{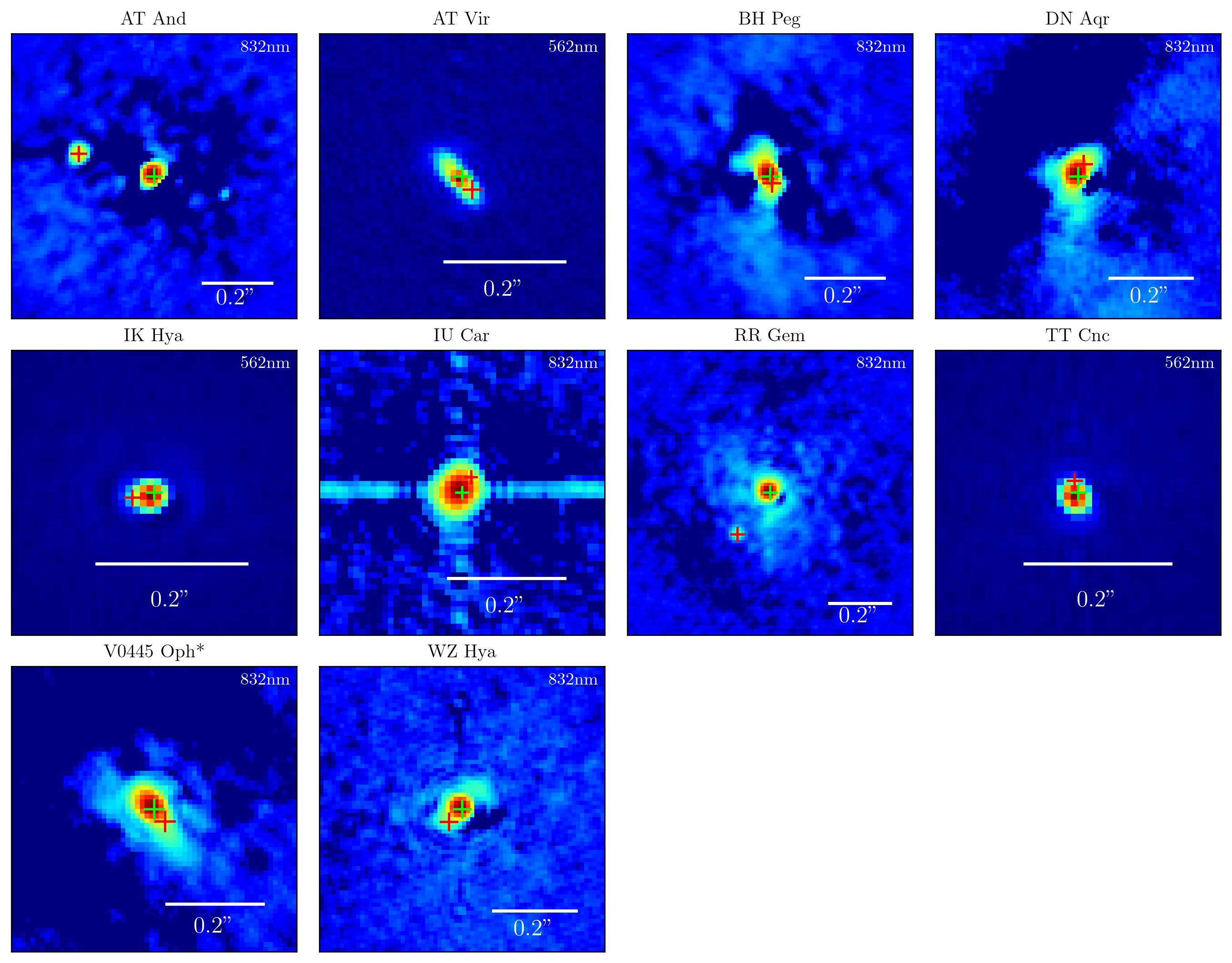}
 \caption{Reconstructed images for the 10 RRL with binary companions. In each plot, the position of the RRL is marked with a green cross, while its companion is depicted in red. The right upper corner indicates which filter is displayed. Each reconstructed image show as well a ``mirror'' image of the companion at 180$\degr$ (more clearly seen in AT And). The real companion is found using bi-spectral analysis \citep{weigelt77}.  }
    \label{fig:reconstructed}
\end{figure*}

\subsection{Optical multi-color photometry}

 Given that the photometric uncertainties inherent to speckle interferometry, especially near the diffraction limit, can be significant, potentially rendering a misidentification about the nature of the RRL companions, we decided to obtain multicolor photometry of several RRL both with and without detected companions. RRL colors at minimum light of their pulsation cycle have been shown to be nearly constant \citep[e.g.][]{blanco92,kanbur96}, and therefore deviations from colors at minimum light can provide insight on the reliability of speckle colors and on the nature of possible companions.

A subset of southern RRL, both with and without detected speckle companions, were observed in the SDSS $griz$ bands with the Zbigniew Ko{\l}aczowski 0.8m telescope (hereafter ZB08) at the Rolf Chini Cerro Murphy Observatory (in short OCM\footnote{\href{https://ocm.camk.edu.pl/}{https://ocm.camk.edu.pl/}}). ZB08 is equipped with a 2k$\times$2k Andor iKon-L 936 camera with a field of view (fov) of 17.2\arcmin$\times$17.2\arcmin. Observations were conducted between March 2025 and January 2026. These observations are part of a larger photometric survey of nearby RRL which will be presented in Zgirski et al. (in preparation), together with the reduction and photometry descriptions which were conducted  with the OCM Fits Processor (OFP, Kicia et al. in preparation).  Photometry was calibrated using secondary standards on each field with calibrated SDSS magnitudes from the Gaia synthetic photometry \citep{montegriffo23}. The targets presented in this paper are discussed in Sect. \ref{sec:minimum}, restricted to $gri$ observations.

\section{Analysis and results} \label{sec:results}

As explained in the previous section, the main results from speckle observations are relative contrast curves in the filters EO562 and EO832 for all targets, and the magnitude difference and separation for the detected companions. In the following subsections we explained how these relative magnitudes are calibrated into absolute magnitudes, and how mass estimations are obtained from both the contrast curves and the detections. We then proceed to discuss each detection individually, as well as some notable non-detections, both in the context of speckle and multi-color broadband observations.

\subsection{Calibration and mass limits for companions}\label{sec:mass_limits}

The 5-$\sigma$ contrast curves from speckle observations in both filters are shown for the complete sample, in Appendix \ref{app:contrast}. Table \ref{tab:contrast} shows as a summary, the 5-$\sigma$ magnitude contrast obtained at distances of 100 mas and 1\arcsec\, in both filters, when available, for all 81 targets. 

These relative magnitude contrast curves need to be calibrated into an absolute scale, and then transformed into mass limits for any possible companions. Exact calibration would require having flux-calibrated spectra of each RRL taken at the same pulsation phase when the speckle observations were taken, which are unavailable. As an approximation to the  RRL spectral energy distribution near minimum light, we used instead two stellar templates from the \citet{pickles98} stellar spectral flux library, a metal-poor F5V, and a  giant F5III star, as the spectral type of RRL near minimum light is known to be early to mid F \citep[e.g.][]{preston59}, together with the \textit{Gaia} $G$ magnitude and distance to each RRL \citep{bailerjones21} to obtain synthetic photometry in the 562 nm and 832 nm speckle filters using the transmission curves provided by Gemini\footnote{\href{https://www.gemini.edu/instrumentation/alopeke-zorro/components}{https://www.gemini.edu/instrumentation/alopeke-zorro/components}}. Calculations were accomplished using \textsc{pyphot}\footnote{\href{https://mfouesneau.github.io/pyphot/}{https://mfouesneau.github.io/pyphot/}}.
Note that the \textit{Gaia} $G$ magnitude used for each target is not the one given directly by \textit{Gaia} DR3 \citep{gaiadr3}, but instead the magnitude expected at the time of speckle observations,  interpolated from the \textit{Gaia} time series photometry using the periods given in Table \ref{tab:sample}.   Table \ref{tab:sample} gives these magnitudes as well as the Julian dates of the mid-point of the speckle observations.

The synthetic 562 nm and 832 nm photometry for each RRL had magnitude differences of at most 0.01 mag when using the two templates, and we adopted the mean of both measurements. The mass limits given in Table \ref{tab:contrast} are then derived by placing each RRL on their respective isochrone, adding the contrast magnitude to the RRL magnitude and finding the isochrone mass associated to that magnitude. The color-magnitude diagrams (CMD) for each RRL were constructed using the MESA isochrones and stellar tracks \citep[MIST,][]{dotter16,choi16} isochrones with metallicities as given in Table \ref{tab:sample} and alpha-element abundances from the relation given by \citet{crestani21} (their Equation 1) using the Stellar Population Interface for Stellar Evolution and Atmospheres \citep[SPISEA,][]{hosek20}, that allows for easier introduction of user-supplied filters. Extinction for the 562 nm and 832 nm observations was calculated following the \citet{cardelli89} law with $E(B-V)$ values taken from the \citet{lallement18} dust maps. Since RRL are traditionally considered an ancient population, an age of 12 Gyr was assumed for all systems (but see Section \ref{sec:thin}).  Changing the age between 10 to 13 Gyr had negligible impact on the mass limits since the contrast limit are always below the main-sequence turnoff magnitude. Fig. \ref{fig:cmds} shows the CMDs for the detected companions, to be discussed in the following section.

Mass upper limits for non-detected physical companions to each RRL are given in Table \ref{tab:contrast}.

\subsection{RRL with detected companions}\label{sec:companions}

\begin{table*}
\caption{Properties of the observed companions}
\label{tab:companions}
\centering
\begin{tabular}{lccrrccccc}
\hline\hline
RR Lyrae   & Filter & Seeing & PA & $\rho$ & $\Delta m$ & $q'$& prob 1& prob 2&  Mass\\
& &  [arcsec] &  [degrees]  & [mas]  & &[arcsec$^2$]& & & [M$_{\odot}$]\\
\hline
AT And & 562 & 0.59 & 75.3 &221 & 2.50 & 0.130 & 2.09E-04 & 2.00E-04&0.84$\pm$0.05\\
       & 832 & 0.50 & 75.7 &222 & 2.37 & 0.111\\
AT Vir & 562 & 0.87&220.1&37&0.0&0.032&1.06E-06&1.00E-05&--\\
       & 832 & 0.79&219.9&28&1.31&0.022\\
BH Peg & 562 & 0.48&196.4&39&2.79&0.019&2.23E-06&4.00E-06& 0.80$\pm$0.05\\
       & 832 & 0.41&196.9&38&1.96&0.016\\
DN Aqr & 832 & 0.48& 335.7& 32 & 1.77 &0.016&9.33E-07& 2.00E-06& 0.79$\pm$0.05 \\
IK Hya & 562 & 0.77&99.5&20&1.73&0.016&1.15E-06 & 4.00E-06& 0.79$\pm$0.05\\
       & 832 & 0.68&102.8&24&1.92&0.016\\
IU Car & 832 & 0.85& 330.1 &31 &1.93 & 0.026&3.43E-06&2.00E-06\\
RR Gem & 562 & 0.67 & 146.6&171 &4.91 & 0.114&8.65E-05&8.20E-05& 0.68$\pm$0.05\\
       & 832 & 0.64 &143.9 &174 & 4.67 & 0.112&\\
TT Cnc & 562 & 0.69& 22.2 & 31 &2.20 &0.022&4.06E-05&3.40E-05& 0.78$\pm$0.05\\
       & 832 & 0.56 & 17.0 & 37 & 2.35 &0.021\\
V0445 Oph & 562 & 0.70&221.6&43&0.45&0.030&5.08E-06&7.40E-05&--\\
          & 832 & 0.60&221.6&27&1.23&0.016\\
WZ Hya & 832 & 0.69&136.0&51&3.41&0.035&3.41E-06&4.40E-06 & 0.76$\pm$0.05\\
\hline
\end{tabular}
%\tablefoot{}
\end{table*}

\begin{figure*}
    \centering
    \begin{subfigure}[t]{0.19\textwidth}
        \centering
        \includegraphics[width=\linewidth]{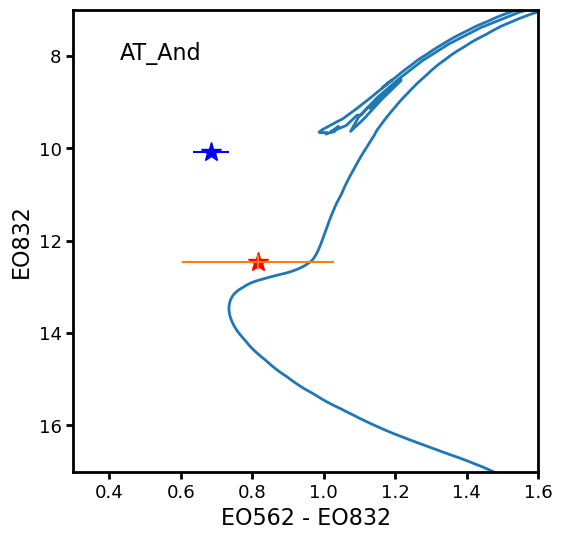} 
    \end{subfigure}
    \begin{subfigure}[t]{0.19\textwidth}
        \centering
        \includegraphics[width=\linewidth]{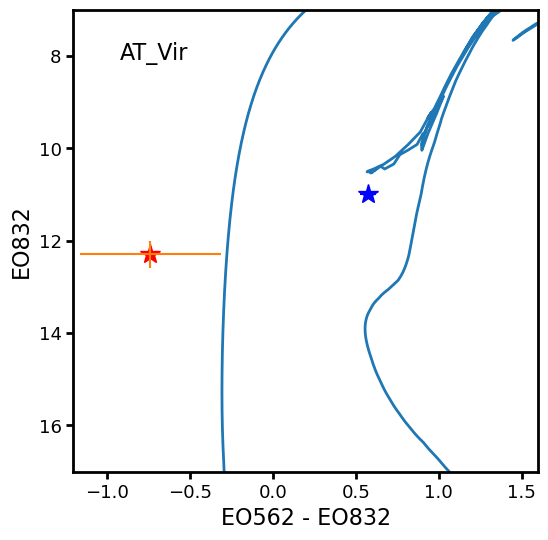} 
    \end{subfigure}
    \begin{subfigure}[t]{0.19\textwidth}
        \centering
        \includegraphics[width=\linewidth]{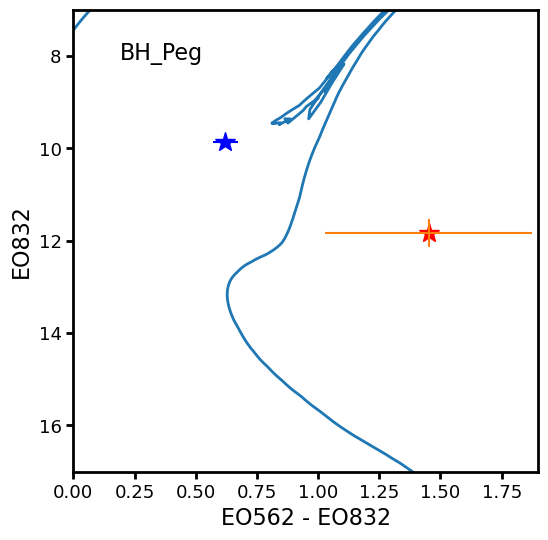} 
    \end{subfigure}
    \begin{subfigure}[t]{0.19\textwidth}
        \centering
        \includegraphics[width=\linewidth]{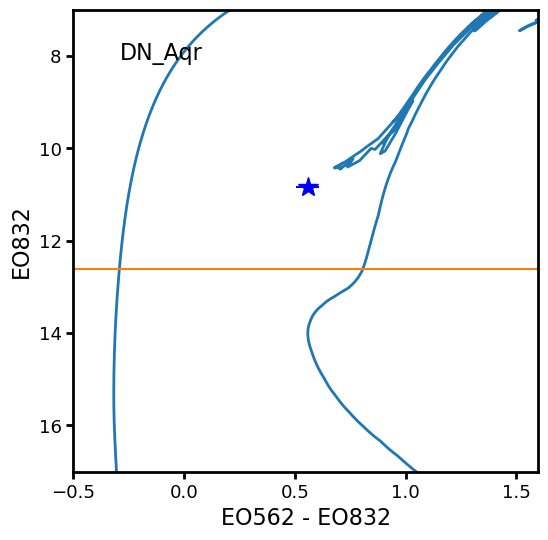} 
    \end{subfigure}
    \begin{subfigure}[t]{0.19\textwidth}
        \centering
        \includegraphics[width=\linewidth]{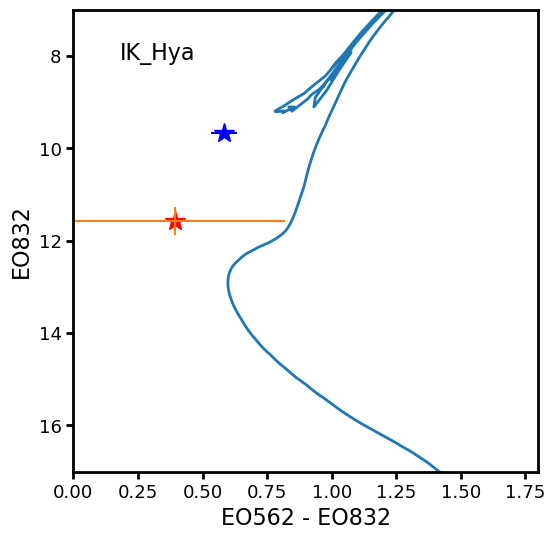} 
    \end{subfigure}
    
    \begin{subfigure}[t]{0.19\textwidth}
        \centering
        \includegraphics[width=\linewidth]{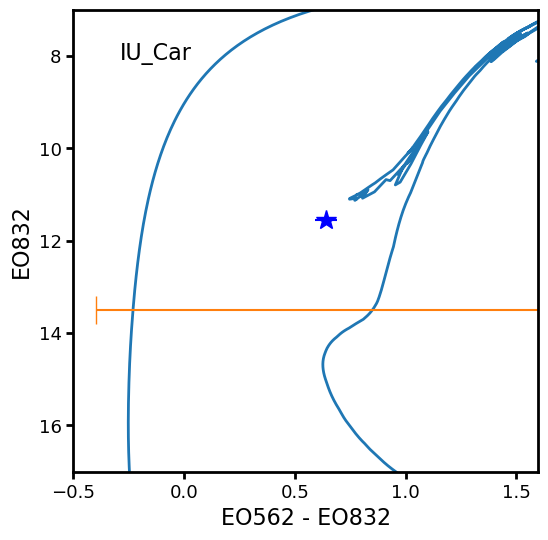} 
    \end{subfigure}
    \begin{subfigure}[t]{0.19\textwidth}
        \centering
        \includegraphics[width=\linewidth]{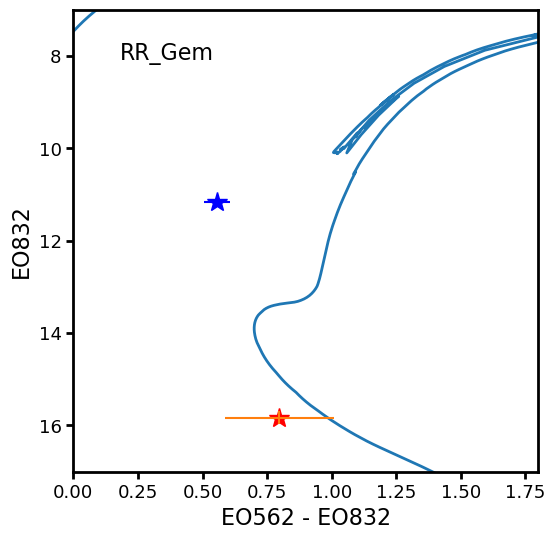} 
    \end{subfigure}
    \begin{subfigure}[t]{0.19\textwidth}
        \centering
        \includegraphics[width=\linewidth]{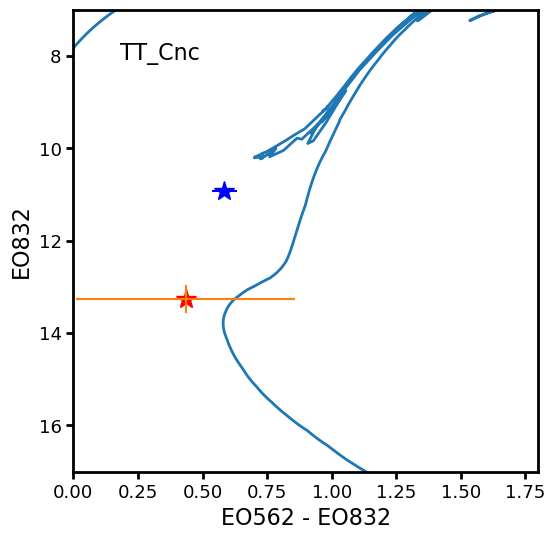} 
    \end{subfigure} 
     \begin{subfigure}[t]{0.19\textwidth}
        \centering
        \includegraphics[width=\linewidth]{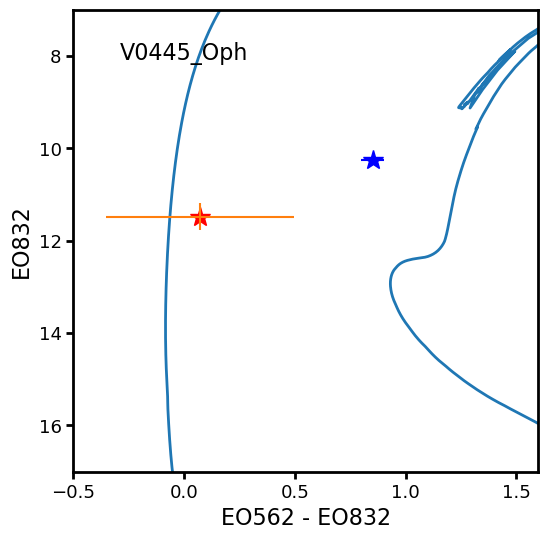} 
    \end{subfigure}
        \begin{subfigure}[t]{0.19\textwidth}
        \centering
        \includegraphics[width=\linewidth]{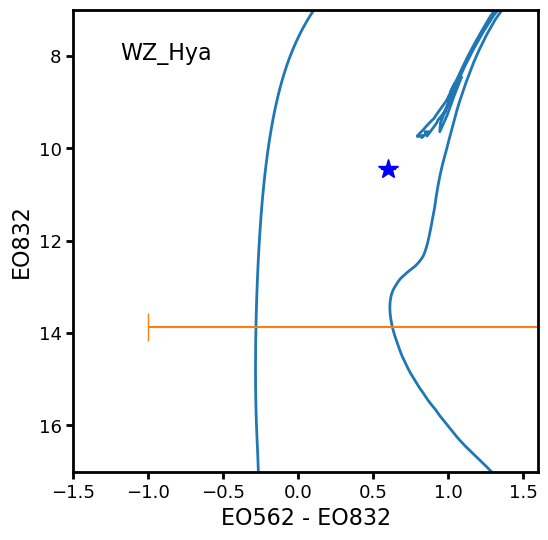} 
    \end{subfigure}    
    \caption{EO562--EO832 vs EO832 color magnitude diagrams for the 10 RRL (blue stars) with their respective companions (orange stars). In light blue are shown  MIST isochrones in the speckle filters with a fixed age of  12 Gyr and the metallicity of each RRL as shown in Table \ref{tab:sample}. In the case of non-detection in the blue camera (DN Aqr, IU Car and WZ Hya), a lower limit for the color is given based on the contrast obtained at that separation from the RRL.}\label{fig:cmds}
\end{figure*}

Companions are detected around 10 out of the 81 RRL observed via speckle imaging. Properties for these companions, including their position angle, the separation between the companion and the target star, $\rho$, as well as its magnitude difference are given in Table \ref{tab:companions}. Additionally, the table provides the seeing in which the observation were conducted as well as $q'$, which is well below 0.6 arcsec$^2$ for all the detections, and finally the companion masses as estimated in Sect.\ref{sec:masses}.

 AT And has the highest PMa among all RRL in the \citet{kervella19a} sample and a very high \textit{Gaia} DR3 renormalized unit weight error \citep[RUWE,][]{lindegren18} of 13.99. RUWE assesses the quality of Gaia's astrometric solution, and high values  are potentially caused by unresolved binaries. \citet{lindegren18} recommends RUWE=1.4 as the threshold between good and bad astrometric solutions. No period changes are detected in the $O-C$ analysis of \citet{leborgne07}, although previously period changes were reported by \citet{olah78}. The PMa analysis of \citet{kervella19a}, under the assumption of a mass of $0.6\,M_\odot$ for the RRL and $0.3\,M_\odot$ for its companion predicts, for a circular orbit, a maximum angular separation of $7^{+2.3}_{-1.9}$ mas, and a period $5.9^{+2.3}_{-1.8}$ years.

Speckle observations reveal a companion to AT And at 222 mas, with a clear detection in both cameras (Fig. \ref{fig:contrast}, upper left panel). Its magnitude and color are consistent with a star at the base of the RGB, or in the sub-giant branch (Fig. \ref{fig:contrast}). 

The \citet{kervella19a} separation prediction of 7 mas is widely different from the 220 mas speckle detection. Changing the parallax used by \citet{kervella19a} from 2.21 (\textit{Gaia} DR2) to the \textit{Gaia} eDR3 value of 1.15 only exacerbates the difference, resulting in a separation of 1.3 mas. Increasing the mass of the companion from 0.3 $M_\odot$, the companion mass assumed by \citet{kervella19a}, to 0.8 $M_\odot$, the mass we obtain from the isochrone, would negligibly ease the tension, predicting a  separation of 10 mas. Given the very low probability that this companion is a chance alignment (see Sect. \ref{sec:chance}), and the impossibility that a companion separated by 220 mas (201 AU for the \citealt{bailerjones21} distance of 915 pc) would produce such a high proper motion anomaly, we put forward the idea that, if the PMa measurement is real, this is a triple system; a first for an RRL. On the other hand, the fact that the $O-C$ analysis of \citet{leborgne07} found a constant period for this star in a light curve spanning 101 years, a much longer baseline than the orbital period proposed by \citet{kervella19a}, puts some doubts on the validity of the PMa measurement.

Finally, while \citet{fernley97} suggested that AT And, partly because of a relatively small light curve amplitude, could be instead an anomalous Cepheid, this smaller amplitude is better explained given the presence of this companion.

AT Vir was included in the \citet{kervella19a} sample who found an inconspicuous SNR of 2.05 for its PMa, considering it non-significant. It also has a \textit{Gaia} RUWE value of 1.26, considered as an indication of a good astrometric solution, and therefore pointing to single stars \citep{lindegren18}. The speckle observations reveal a very close companion, with some discrepancy in the separation found by the two cameras (37 mas in the blue channel, 28 mas in the red channel), although these are expected at separations close to the diffraction limit \citep[][and see Sect.\ref{sec:obs}]{horch11}. 

The speckle observations for this target in EO562, with a $\Delta$m=0, reveals a detection but with a very poor fringe fitting. Taken at face value, the speckle CMD reveals what appears to be a very blue companion, marginally consistent with a hot post-AGB star or a hot subdwarf (see respective panel in Fig. \ref{fig:cmds}). The nature of this companion, and the reliability of its speckle color, are discussed with the aid of multicolor photometry in Sect. \ref{sec:minimum}.

 BH Peg, included also in \citet{kervella19a}, but like AT Vir, its PMa was considered non-significant, and similarly it also has a low RUWE (0.89). \citet{leborgne07} found that the residuals over a parabolic fit of an $O-C$ diagram of its light curve could be interpreted as evidence of binarity. Speckle observations reveal again a very close companion at 38 mas, detected in both cameras. This time, even though its EO832 magnitude is consistent with the base of the RGB, its colour appears significantly redder (see top left panel in Fig. \ref{fig:cmds}), consistent with a ``red straggler'' \citep[e.g.][]{shishkovsky18}, although again the proximity of the detection to the diffraction limit makes its photometry quite uncertain. 

 DN Aqr also presents inconspicuous PMa and RUWE. It was observed when Zorro's blue camera was misaligned, and therefore there is no colour information  about its detected companion, which appears on the red camera at a separation of 32 mas. Fig. \ref{fig:cmds} indicates only the EO832 magnitude level of the companion, and gives the full range in EO562--EO832 as possible colour. 

IK Hya is among the RRL with the lowest PMa in the \citet{kervella19a} sample. It shows a large modulation in its amplitude \citep{leborgne12} with a complex set of frequencies \citep{skarka14} as also visible by the line doubling of Balmer lines in its spectrum \citep{kalari24b}. The speckle observations show a companion detected in both cameras at an average separation of 22 mas, the closest detection among the sample. Its (EO562--EO832) color is consistent with a blue straggler, with the usual caveats for such close detections.

IU Car has no indications of binarity in the literature (RUWE=1.17), with perhaps its only peculiarity being the existence of a period jump \citep{szczygiel07}. The companion is only detected in the red camera of Zorro at a separation of 31 mas. The blue contrast curve at the same separation very loosely constrains the nature of this companion which can have any colour (EO562--EO832)>--0.4. 

 RR Gem is a well-known Bla\v{z}ko pulsator, with the shortest Bla\v{z}ko period among RRab, and a very small modulation amplitude \citep{jurcsik05,sodor07}. It has no evidence of companions from PMa or \textit{Gaia} RUWE, nor from a dedicated RV search \citep{barnes21}. The older study of \citet{firmanyuk76}, based on $O-C$ analysis, found a periodicity of 25\,600 days ($\sim$70 years). Speckle observations show a companion in the upper main sequence. With a $\Delta$EO832=4.67, this is the faintest companion detected in this study. Its color and magnitude are consistent with a MS star.
   
 TT Cnc was studied by \citet{kervella19a} who found a low signal-to-noise ratio for its PMa and therefore did not considered it as part of a binary system. \textit{Gaia} DR3 provides an inconspicuous RUWE value of 1.18. A companion is detected in both cameras at 34 mas on average, with a color consistent with the MS turn-off of its respective isochrone. 

V0445 Oph is another RRL with no previous suspicion of binarity, with a normal RUWE of 1.05 and no significant PMa.  The speckle detection, in both cameras, is again close to the diffraction limit, at an average separation of 35 mas. Like in the case of AT Vir, its speckle color (EO562--EO832) is very blue and consistent with a subdwarf or similarly hot star. With a metallicity of [Fe/H]=--0.23, it is the most metal rich star with a detected companion in our sample (see Sect. \ref{sec:thin}).

 WZ Hya had no indication of binarity, with both a low RUWE and PMa. The speckle observations detected a companion at 51 mas, albeit visible only in the red camera. The blue contrast curve at that separation provides a loose color constraint for the companion of EO562-EO832>--1.0.

\subsection{Notable RRL with no detected companions}\label{sec:no_detect}

 UV Oct, with a high PMa only surpassed by AT And (see Sect. \ref{sec:companions}), and with a predicted companion at 0.6$^{+0.4}_{-0.2}$\arcsec with a mass of 0.3$\pm$0.18\,M$_{\odot}$ \citep{kervella19a}, was one of the prime targets on this search for RRL companions. It was studied by \citet{salinas20}, reaching a contrast of 4.8 mag at 0.6\arcsec, corresponding to main sequence mass of 0.67 M$_{\odot}$, and therefore unable to reach the mass range predicted by \citet{kervella19a}. Our new observations, doubling the previous exposure time, reach a contrast in the EO832 filter of 7.52 mag, ruling out a companion mass above 0.37 M$_{\odot}$ at 0.6\arcsec\, based on our CMD analysis (Sect. \ref{sec:mass_limits}). At 1\arcsec, the reached contrast of 8.14 mag, rules out a mass above 0.28 M$_{\odot}$. 

TU UMa is the poster child of RRL in binary systems, as the only one with a spectroscopically confirmed companion. Evidence for this companion comes mostly from the discovery and analysis of cycles in its $O-C$ diagram \citep[e.g.][]{szeidl86,kiss95} as well as radial velocities \citep{saha90}. The latest analysis of its light-time travel effect is that of \citet{liska16tu} who found a minimum mass for the companion of 0.33 M$_{\odot}$ with an orbital period of 23.3 yr in a significant elliptical orbit with major axis close to 3 AU.  \citet{kervella19a}, on the other hand, proposed that the companion would be a compact remnant with a mass close to 2 M$_{\odot}$ with a separation of 11 AU. At the distance of TU UMa (645 pc), speckle interferometry cannot image directly the  proposed separations since the diffraction limit at this distance is 12.9 AU, but it can search for a possible third star in the system. With these observations, we rule out an external main sequence companion with mass higher than 0.69 M$_{\odot}$ and 0.18 M$_{\odot}$ at separations of 0.1\arcsec\,and 1\arcsec, respectively.

BB Vir was found to have anomalous color and spectra by \citet{kinman92}, who concluded that these observations could be reconciled if BB Vir was in a binary system with a blue horizontal branch (BHB) star. The anomalously high flux in the blue was confirmed by ultraviolet spectra obtained by \citet{fernley93}, also supporting the idea of a blue companion. The alternative explanation, that BB Vir is simply a hot RRL very close to the blue edge of the instability strip, is inconsistent with the shape of its light curve, undistinguishable from an RRab type, and its low amplitude \citep[$A_V=0.63$, in ASAS-SN,][]{shappee14}, which should be higher than usual for a hot RRab \citep{kinman92}.  The $O-C$ analysis of \citet{liska16lite} also supports the  binary system hypothesis, with their best fitting model yielding a period of 93 yr. Finally, the assertion from \citet{muraveva18} that ``BB Vir turned out to be a blend of two stars'' appears without any explanation and it seems unfounded given the absence of near companions in both our speckle observations (see below) and \textit{Gaia}, albeit \textit{Gaia}'s RUWE is significantly high, with a value of 4.26.

A BHB companion may have a very similar brightness compared to an RRL, especially those cooler than about 12\,000 K \citep[e.g.][]{catelan09}, and therefore should be detected down to diffraction limit ($\sim$20 mas) of these speckle instruments. The non-detection therefore puts constraints on the nature of its putative companion. At the distance of BB Vir (1.7 kpc), 20 mas means a separation of 34 AU. Using Kepler's 3rd law, we can estimate the separation of BB Vir and its putative BHB companion using the 93 yr period from \citet{liska16tu}, and further assuming masses of 0.5 M$_{\odot}$ for both the RRL and the BHB star \citep{heber97} obtaining a separation of 94.4 AU or 55 mas, where the observed contrast in the EO562 filter is around 2 mag. The lack of detection at this separation therefore puts any possible companion into the realm of hot BHBs (T $\gtrsim$ 12\,000 K) or even in that of extreme horizontal branch (EHB) stars, where potentially $u$-band speckle observations could yield a direct detection. We further discuss this target in the context of minimum light colors in Sect. \ref{sec:minimum}.

\subsection{Chance alignments}\label{sec:chance}

Despite all RRL with detected companions having fairly high Galactic latitudes, with the lowest being AT And with $b=-18\degr$, we evaluate the probability that their detected companions are simple chance alignments instead of bound companions. We calculate this probability in two different ways, both using \textit{Gaia} DR3 detections around each RRL.

The first method assumes that the distribution of stars around each RRL is essentially random, following a uniform distribution. Under this assumption, the probability of a star being closer than an angular separation $\rho$ to a particular target is given by $P=1-e^{-\pi\Sigma^2\rho}$, where $\Sigma$ is the surface density of objects in the target's neighbourhood \citep[e.g.][]{correia06}. $\Sigma$  is calculated by taking all \textit{Gaia} DR3 sources within 5\arcmin\, from each RRL with a detected companion. The resulting probabilities for each companion being a chance alignment are listed in Table \ref{tab:companions}, in the ``prob 1'' column. 

The second method is that of \citet{neugent20}, where the position of the target with the binary companion is moved randomly within its environment, and the number of times in which a star in the environment falls within a distance of $\rho$ is counted to form a probability. This bootstrapping approach is repeated 500\,000 times for each target using as environment again all \textit{Gaia} DR3 sources within 5\arcmin. Chance alignment probabilities from this method are given in the ``prob 2'' column in Table \ref{tab:companions}.

Regardless of the method, the probabilities of chance alignments are very low, with AT And, the one with the lowest Galactic latitude, and the farthest detected companion at 220 mas, having the highest probability of chance alignment with a mere 0.2\%.  It is worth noting that the real probabilities of chance alignments are even lower, since we did not consider either the magnitude or the colour constraints imposed by the speckle detections, and instead we took all \textit{Gaia} sources in each RRL neighbourhood down to \textit{Gaia}'s limiting magnitude. 

The results of these two approaches give us confidence in stating that all speckle detections are very likely gravitationally bound companions, despite the photometric uncertainties inherent to speckle interferometry, especially when close to the diffraction limit.

\subsection{The companion masses}\label{sec:masses}

 The extremely low probability that any of the speckle detections is a chance alignment puts strong constraints on the nature of the detected RRL companions since all of them should share the same properties (distance, age, metallicity, reddening) of their RRL, and therefore lie on the same isochrone. This allows us to assign masses to the companions, comparing their speckle magnitudes with their respective isochrone. Companions masses, just like mass limits in Sect. \ref{sec:mass_limits}, are therefore simply estimated from the CMDs (see Fig. \ref{fig:cmds}), finding the closest point from the companion color and magnitude and its respective isochrone. For those that only have a measurement in EO832, the mass comes from an interpolation of their EO832 detection level and the isochrone. Masses of each companion are given in Table \ref{tab:companions}.

Uncertainties for the companion masses come from uncertainties in the distances, metallicities, ages, but the dominant factor is the uncertainty in the speckle photometry itself, which we have assumed as 0.1 mag for companions father out than 50 mas, and 0.3 mag for those inside this limit. Despite this large photometric uncertainty, since all companions are restricted between the upper main sequence and the low RGB, the impact on their masses is small, less than 0.1 M$_{\odot}$.
The very blue colors for the companions of AT Vir and V445 Oph as seen from their respective CMDs are in conflict with the multicolor photometry presented in Sect.\ref{sec:minimum} and therefore we refrain from deriving their masses following the same approach.

The mean mass of the companions is 0.77$\pm$0.04 M$_{\odot}$. This is broadly consistent with the high mass peak of the mass distribution of companions found by \citet{hajdu21} of $\sim$0.6 M$_{\odot}$ which they associate with a mix of white dwarves and main-sequence companions.

\subsection{Common proper motion pairs}

An additional way to search for RRL with wide companions, lying beyond the small fov where speckle interferometry is applicable, is by using \textit{Gaia} astrometry. Pairs of stars sharing the same parallax as well as the same proper motion can be considered as genuine physical wide binaries \citep[e.g.][]{fouesneau19}. To search for RRL in these common proper motion (CPM) pairs, we adopt the approach of \citet{pelisoli20} who searched for CPM pairs among hot subdwarfs. Briefly, we queried \textit{Gaia} DR3 using an ADQL search to select pairs consistent within 3$\sigma$ in both parallax and proper motion. We further applied the quality cuts recommended by \citet{lindegren18}, that is,

\begin{equation}
1.0 + 0.015\,(G_{\rm BP} - G_{\rm RP})^2 < E < 1.3 + 0.06\,(G_{\rm BP} - G_{\rm RP})^2,
\end{equation}
where $E$ is the  photometric excess factor, \texttt{phot\_bp\_rp\_excess\_factor}, and

\begin{equation}
u < 1.2 \max (1, \exp[-0.2(G - 19.5)]),
\end{equation}
with $u = \sqrt{\frac{\rm astrometric\_chi2\_al}{\rm astrometric\_n\_good\_obs\_al -5}}$, where \texttt{astrometric\_chi2\_al} and \texttt{astrometric\_n\_good\_obs\_al} correspond to the astrometric chi-square statistic and the number of valid along-scan observations, respectively, as provided in \textit{Gaia} DR3. We additionally imposed that the uncertainties in both parallaxes and proper motion remain below 15\%.

After following this procedure, we find no CPM pairs in \textit{Gaia} DR3 for the 81 RRL observed with speckle interferometry.

We also revisit the CPM pairs found by \citet{kervella19b} who used \textit{Gaia} DR2 astrometry determining 7 CPM pairs among 789 RRL they studied. Using the same approach described above, we find that only one (RR Leo) out of this seven RRL can still be considered as a CPM pair under \textit{Gaia} DR3 astrometry (see Table \ref{tab:cpm}). RRL OV And deserves a special mention, as it was considered by \citet{kervella19b} as being in a CPM pair with an F4V star. The pairing of an old RRL with a relatively young, massive MS star should not be possible, with the exception being if the companion was a blue straggler, or in the more exciting case of a ``young RRL'' (see Sect. \ref{sec:thin}). Unfortunately, the explanation is far more trivial, and we find that \citet{kervella19b} misidentified OV And in the \textit{Gaia} catalogue; while OV And has \textit{Gaia} ID 380489851379496448, \citet{kervella19b} used instead 380489851379496320, which corresponds to a nearby star, $\sim$7\arcsec away.

In summary, our search for CPM pairs shows that the existence  of companions in very wide orbits ($\gtrsim$ 500 AU) seems extremely rare for RRL, which goes in line with the results of \citet{lodieu25}, who found a binary fraction of at most 3\% for solar-mass metal-poor ([Fe/H]<--1.5) stars between 8 and 10\,000 AU.

\subsection{Minimum light colors and the nature of close companions}\label{sec:minimum}

As explained in Sect. \ref{sec:speckle} and as shown in the CMDs in Fig. \ref{fig:cmds}, the speckle photometry, when companions approach the diffraction limit, becomes quite uncertain, particularly affecting what can be learned from the speckle colors.  In order to better assess the speckle color information, we obtained multicolor photometry for a subset of RRL, particularly aiming at measuring their colors during minimum light. The study of minimum light has two advantages. First, the contrast between the RRL and its companion is minimized, making the impact of the companion in the compound colors more clear; and second, the colors of RRL at minimum light are a well-established ``standard crayon'', that is, they have a mostly constant value with a very mild dependency with the stellar parameters of the RRL \citep[e.g.][]{blanco92}, and therefore any anomalous colors at this pulsation phase give us a glimpse into the nature of the binary companions.

We obtained $griz$ light curves for AT Vir, BH Peg, DN Aqr, IK Hya, IU Car, and WZ Hya, all of which present speckle companions (see Sect. \ref{sec:companions}); to these we added BB Vir, for which we do not detect a speckle companion, despite a long-suspected blue companion (see Sect. \ref{sec:no_detect}). As a control sample, we obtained $griz$ light curves for RX Eri, RZ Cet, SV Eri, SZ Gem and WY Ant, sharing a similar range of metallicities and periods to the RRL with companions, and also presenting low reddenings ($E(B-V)<0.1$).  None of the stars in the control sample reveal the presence of companions by our speckle observations or other methods, and therefore represent what is expected for normal minimum light colors. 

The definition of minimum light changes across the literature. One common approach is to take the average color between phases 0.5 and 0.8 \citep{sturch66}, while \citet{kunder10} adopted the $V-R$ color when the $V$ light curve has a minimum, and \citet{vivas17} simply took phase 0.65 after template fitting. In our case, given the excellent sampling of the light curves, we take the minimum light (or maximum magnitude) itself at each band after a  cubic spline  smoothing \citep{deboor78}\footnote{\href{https://github.com/espdev/csaps}{https://github.com/espdev/csaps}},  which followed more accurately the shape of the curves near minimum light than classical Fourier series, with the exception of IK Hya (see below). Two examples of the $griz$ light curves are shown in Fig. \ref{fig:lcs}.

\begin{figure}[t]
    \centering
    \includegraphics[width=\columnwidth]{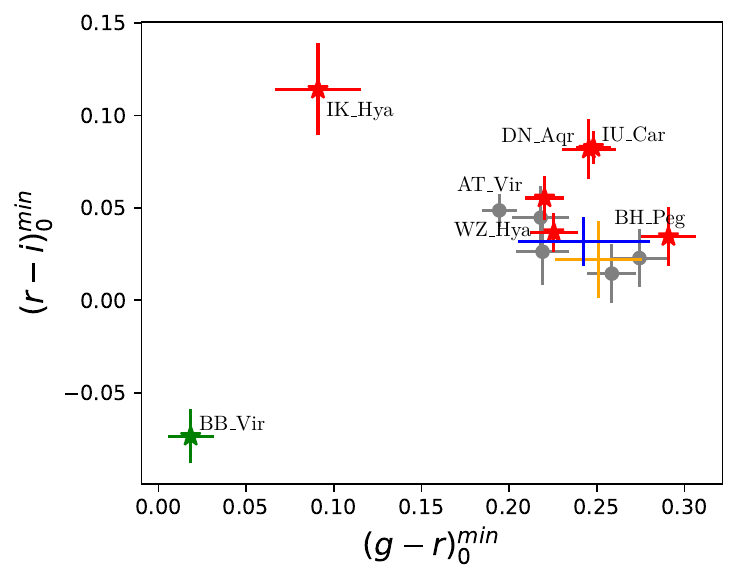}
    \caption{Color-color plane for a sample of RRL with speckle companions (in red stars) and a control sample of RRL without companions (grey dots). BB Vir is indicated in green. The large blue cross shows the mean and 1-sigma dispersion for $(g-r)_0^{\rm min}$ and $(r-i)_0^{\rm min}$ for our control sample, while the orange one shows the same for RRab stars in M 5 from \citet{vivas17}.}
    \label{fig:colors}
\end{figure}

Fig. \ref{fig:colors} shows the color-color plot of $(g-r)_0^{\rm min}$ versus $(r-i)_0^{\rm min}$ of both the RRL with detected companions (in red symbols) and those of the control sample (in grey symbols). Colors were dereddened using the \citet{schlafly11} dust maps and the \citet{fitzpatrick99} reddening law with $R_V$=3.1. The control sample concentrates around $(g-r)_0^{\rm min}$=0.242$\pm$0.037 and  $(r-i)_0^{\rm min}$=0.031$\pm$0.013; this can be compared to the values found by \citet{vivas17} in the globular cluster M 5, $(g-r)_0^{\rm min}=0.251\pm0.025$ and $(r-i)_0^{\rm min}=0.022\pm0.021$; remarkably close considering that the definitions for minimum light are different, and that our control sample has a larger range of metallicities and reddenings. The uncertainties of each color come from the photometry and the standardization procedure of each band, while uncertainties in the extinction are not introduced. These colors shed new light into the speckle results described in Sect. \ref{sec:companions}.

Ik Hya is a complex Bla\v{z}ko pulsator (see Sect. \ref{sec:companions}), with a main Bla\v{z}ko period of about 72 days \citep{skarka14}. In order to determine the phase of minimum light we used the $g$ observations to select a range of 15 nights where the majority of observations were obtained; this phase of minimum light in $g$ was adopted to determine the minimum light in the other filters. The minimum light colors are difficult to interpret; while a bluer $(g-r)_0^{\rm min}$ is consistent with the blue companion found with the speckle observation, the redder $(r-i)_0^{\rm min}$ would instead imply a red companion. Even though IK Hya has the largest reddening of all the RRL considered in this section, the inconsistency between the colors can likely be attributed to its Bla\v{z}ko nature. The claim from \citet{guldenschuh05} that minimum light colors remain constant during a Bla\v{z}ko cycle, based on the study of a single Bla\v{z}ko RRL, needs to be revisited with a larger sample in the future. 

In Sect. \ref{sec:companions} it  was shown that AT Vir has a very blue EO562--EO832 color, but its minimum light colors are basically indistinguishable from the control sample. It is therefore very likely that the uncertainty in EO562 is significantly greater than the assumed value of 0.3 mag, and the companion is much closer to the lower RGB than what the speckle colors suggest. 

DN Aqr and IU Car have virtually identical minimum light colors, and also their companions have almost identical contrasts with $\Delta_{EO832}$=1.77 and 1.96, respectively, placing their companions at the base level of the RGB, but with very poor constraint on their color given the non-detection in EO562 in the case of IU Car, and because DN Aqr observations were only taken with the EO832 filter (see respective CMDs in Fig\ref{fig:cmds}). Both RRL have minimum light $(g-r)_0^{\rm min}$ consistent with the control sample, but $(r-i)_0^{\rm min}$ is 3.8$\sigma$ higher (redder) than the control sample. This confirms that these companions are redder than the RRL itself, and likely at the base of the RGB, considering also their very close separation.

BB Vir has a long suspected blue companion, that we do not detect in our speckle observations, but based on that non-detection we suggest it could be a hot BHB or an EHB star (see discussion in Sect. \ref{sec:no_detect}). Its $(g-r)_0^{\rm min}$ and $(r-i)_0^{\rm min}$ colors are remarkably blue, at the 7$\sigma$ and 8$\sigma$ level compared to the control sample, respectively. These very blue colors support the idea that the companion is much hotter than BB Vir, and probably is a hot BHB or EHB star.

Finally, WZ Hya appears with indistinct colors to those in the control sample, showing one of the limitations of the method of minimum light colors to identify companions. The WZ Hya companion is 3.41 magnitudes fainter in EO832 than WZ Hya. The absence of  anomalous colors reaffirms the conclusion that the companion is likely a main-sequence star, since a very blue star, something still allowed by the contrast obtained in the EO562 filter, would have been probably visible in $(g-r)_0^{\rm min}$.

As mentioned above, minimum light colors are not exactly the same for all RRL, but have a small dependency on metallicity and pulsation period. The impact of metallicity in the minimum light colors in the SDSS bands have been studied by \citet{vivas17}, finding that metallicity affects mostly those colors involving the $u$ band, and partly those involving $g$ due to the effect of line blanketing at these wavelengths. The impact on redder bandpasses is almost negligible. For example, the color $(r-i)_0^{\rm min}$ is expected to change only by 0.006 mag in the --2.0<[Fe/H]<--1.0 range. The impact on $(g-r)_0^{\rm min}$ is not explicitly given, but from their Fig. 5, it can be gleaned that is less than 0.01 in the same metallicity range. The anomalous colors in Fig.\ref{fig:colors} cannot be attributed to metallicity differences.

Another effect is the correlation between the minimum light colors and the pulsation period. \citet{vivas17} studied this correlation analyzing 47 RRab in M 5. Again, the impact is very minor for $(r-i)_0^{\rm min}$, and \citet{vivas17} estimates it less than 0.016 mag. In $(g-r)_0^{\rm min}$, the shift introduced in the period range of the RRL considered in this section can be as large as 0.05 mag. Even though we are not explicitly correcting for this shift, the fact that our control sample and the RRL with companions share the same period range implies that anomalous colors do not stem from period differences.

The systematic study of minimum light color in multiple bands reveals as a useful tool for the potential discovery and study of RRL in binary systems, and it can be particularly useful in RRL too faint for speckle interferometry and RV studies. But not without its limitations; minimum light colors can only detect companions which are bright enough ($\Delta$mag $\lesssim$ 2 compared to the minimum light of the RRL), and that have colors different enough from the RRL. Ultimately, its usefulness may depend on how accurate is the information about the reddening to each star, since an incorrect reddening would smear their positions in the color-color plane. This hurdle may be overcome with the usage of reddening-free pseudo-colors \citep[e.g.][]{catelan13}, which have been scantly used in the study of variable stars.

\section{Discussion}\label{sec:discussion}

\subsection{Binary detection rates across methods}

Our speckle search has detected 10 companions among the 81 observed RRL, a 12.3\% detection rate. This detection rate can be compared to other methods used to search for RRL in binary systems. \citet{hajdu15} analyzed the OGLE-III light curves of 1\,952 RRL in the Galactic bulge via $O-C$ diagrams finding 20 binary candidates (from which parameters could be derived for only 12), giving a detection rate of 1\%. The expanded sample of \citet{hajdu21} analzyed 27\,480 $O-C$ diagrams from homogenized photometry from OGLE-III and OGLE-IV of bulge RRL with the discovery 87 RRL in candidate binary systems for a discovery rate of less than 1\%. Similarly, \citet{prudil19} presented 20 RRL in binary systems based on the analysis of 9\,000 $O-C$ diagrams also from OGLE and KMTNet \citep{kim16} photometry again for a discovery rate below 1\%.

\citet{kervella19a} tested for binarity among RRL using the PMa between Hipparcos and \textit{Gaia} DR2 astrometry. From their sample of 198 RRL they find a significant PMa in 13 of them, implying a discovery rate of 6.8\%. 

Other direct or serendipitous methods to find RRL in binary systems include radial velocity measurements, which have not found additional binary candidates \citep{guggenberger16,barnes21};  anomalous chemical abundances \citep[like in TY Gru,][]{preston06,dorazi25}, or anomalous colors (like in BB Vir, discussed in Sect\ref{sec:no_detect}). With the exception of radial velocity measurements these have not been attempted systematically for a significant sample and their detection rates cannot be determined.

At face value, speckle interferometry significantly outperforms other methods solely in terms of the detection of RRL in binary systems, which is not entirely surprising since RRL in binary systems are only expected if the orbital geometry did not allow for any mass transfer, which precludes searches sensitive mostly to close orbits, i.e. radial velocities, or $O-C$ analyses based on short timescales.  Even though speckle interferometry is sensitive to these wide orbits, a small amount of them, close to be edge-on, can remain undetected.

\subsection{Binary fraction}
\begin{figure}
    \centering
    \includegraphics[width=\linewidth]{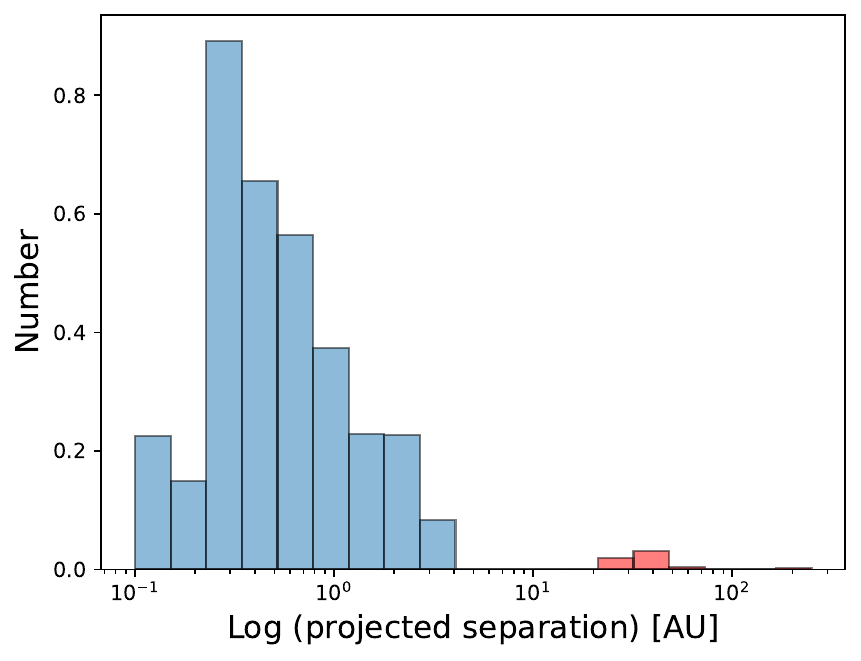}
    \caption{Distribution of the separations of speckle companions (in red), compared to those found via $O-C$ analysis in \citet{hajdu21} (in blue).}
    \label{fig:sep}
\end{figure}

RRL progenitors are very old ($\gtrsim$ 10 Gyr), predominantly metal poor ([Fe/H]<--1.0), main sequence stars with masses slightly below solar; as such, RRL represent one of the oldest stellar populations in the Universe.

Even though significant work has been put in understanding the binary fraction among metal-poor stars \citep[e.g.][]{latham02,raghavan10,moe19,elbadry19}, it is unclear whether the same binary fractions are to be expected for RRL, especially since most studies are restricted to close binaries, which are especially lacking among RRL, for example, \citet{hajdu21} via $O-C$ analysis found no binary candidates below orbital periods of 1\,000 days, and a dearth between 1\,000 and 2\,000 days compared to those with periods higher than 2\,000 days.

Following \citet{kalari24a}, we adopt the Clopper-Pearson approach using binomial statistics, to estimate the confidence intervals for the measured binary fraction of 12.3\%, finding that a binary fraction higher than 25\% can be ruled out at the 99\% confidence level. These numbers are valid within the  projected separation range that speckle interferometry can access, between about 20 AU and 1200 AU for the median distance of our RRL sample of 1120 pc. 

\citet{moe19} found that for close binaries (<10 AU) at [Fe/H]=--1.0, the binary fraction is 40 $\pm$ 6\%, while for wide binaries(>200 AU), the binary fraction is independent of metallicity with a value of 21$\pm$3\%. With a median separation of 34 AU all of our detected companions lie in the \textit{intermediate} region (see Fig. \ref{fig:sep}) defined by \citet{moe19} between 10 and 200 AU for which they simply linearly interpolate the results found between the other two regions, for a binary fraction close to 30\%. So despite obtaining a higher binary fraction than in previous studies, the RRL binary fraction still falls short when compared to the stellar population that is more akin to its progenitors, i.e. metal-poor solar-type stars. But RRL are not alone in this low binary fraction; recently \citet{culpan25}, based on radial velocity measurements, also claimed an alarming lack of binaries among BHB stars, which, just like RRL, are evolved core-helium burning stars. 

\subsection{Thin disc RRL} \label{sec:thin}

\begin{figure}[t!]
    \centering
    \includegraphics[width=\linewidth]{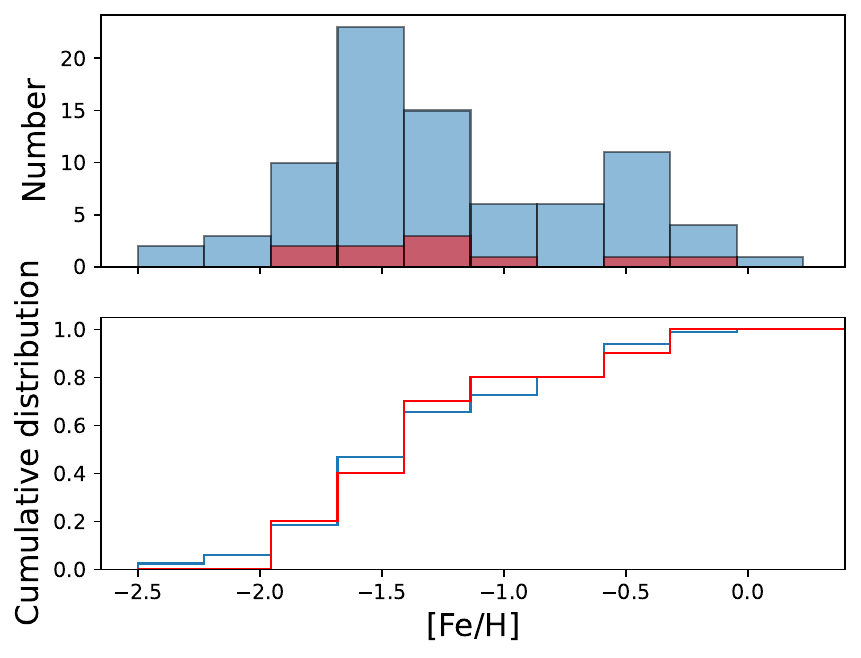}
    \caption{Metallicity distribution of the 81 observed targets, shown as a histogram (upper panel) and a cumulative distribution (lower panel). In both panels the targets without speckle companions are shown in blue, while those with companions are shown in red.}
    \label{fig:metal}
\end{figure}

6-D kinematical information of Galactic RRL from \textit{Gaia} has revealed that a significant number of them share their kinematics with the thin disc. Initially found among the solar neighbourhood RRL \citep{zinn20,prudil20}, the same behaviour has been detected throughout the entire extent of the thin disc \citep{iorio21}. 

The existence of thin disc RRL is puzzling, since, while the age of the thin disc has been constrained to be younger than about 8 Gyr, from both the luminosity function of thin disc white dwarfs \citep[e.g.][]{kilic17}, as well as nucleocosmochronology measurements \citep{delpeloso05}; RRL are considered significantly older, with ages above 10 Gyr. A few possibilities have been raised to resolve this discrepancy. The first possibility is that a metal-rich, old, accreted galaxy, as massive as the LMC \citep{zinn20} was dragged into the disc \citep{feuillet22}, while a second possibility to populate the thin disc with old stars would be radial migration from the Galactic bulge \citep{beraldo21}. A third scenario is that these RRL come from an hitherto unknown old thin disc population \citep{dorazi24}.

A fourth and most tantalizing explanation for thin-disc RRL is the possibility that these may be formed via binary evolution \citep{pietrzynski12,karczmarek17}. \citet{bobrick24} proposed that close binaries can more efficiently strip the envelopes of red giant branch stars, therefore producing horizontal branch stars at younger ages (and therefore from more massive progenitors), ending up inside the instability strip, and therefore becoming ``young RRL''.  According to \citet{bobrick24}, the orbital periods of these systems would lie preferentially between 1\,000 and 2\,000 days.

Our sample contains 13 RRL out of the 22 that share thin disc kinematics according to \citet{prudil20}.  \citet{abdollahi25} also proposed 9 RRL with thin disc kinematics, although six of them were already proposed by \citet{prudil20}. All nine were observed as part of our speckle program. Out of the 16 individual thin disc RRL from the combined samples of \citet{prudil20} and \citet{abdollahi25} observed with speckle, we detect a companion in only one of them, V0445 Oph, putting the measured binary fraction for this subsample at 6.2\%. 

Even though these speckle observations cannot image directly the separations where the predicted companions would be, they show that thin disc RRL have in general a lower multiplicity than the halo RRL population, corresponding to the general finding that metal-rich solar type stars have a binary fraction three times lower than their metal-poor counterparts when considering separations between 50 and a 100 AU \citep{elbadry19}. This lower binary fraction in metal-rich RRL, coupled with the dearth of RRL in very close orbits \citep[][and see Fig. \ref{fig:sep}]{hajdu21}  cast doubts on the idea of a binary origin as a dominant mechanism of RRL formation as put forward by \citep{bobrick24}, but larger samples are needed to confirm this.

\subsection{Metallicity distribution of RRL in binaries}

The metallicity distribution of the 81 RRL in the sample is shown in Fig. \ref{fig:metal}. The metallicity distribution shows the well-known two peaks near [Fe/H]=--1.5, corresponding to the MW halo, and [Fe/H]=--0.5, corresponding to the disc population \citep[e.g.][]{liu20,crestani21}, showing that our sample is not biased to any particular metallicity. The plot also shows the metallicity of those RRL with detected companions via speckle in red.

The cumulative distributions (lower panel in Fig. \ref{fig:metal}) show no significant differences, perhaps only revealing a lack of binaries in the lowest metallicity bin below [Fe/H]=--2.0. A Kolmogorov-Smirnov test gives a high $p$-value of 0.92, suggesting both samples come from the same distribution.

\subsection{Binarity and period luminosity relations}

\begin{figure}[t]
    \centering
    \includegraphics[width=\linewidth]{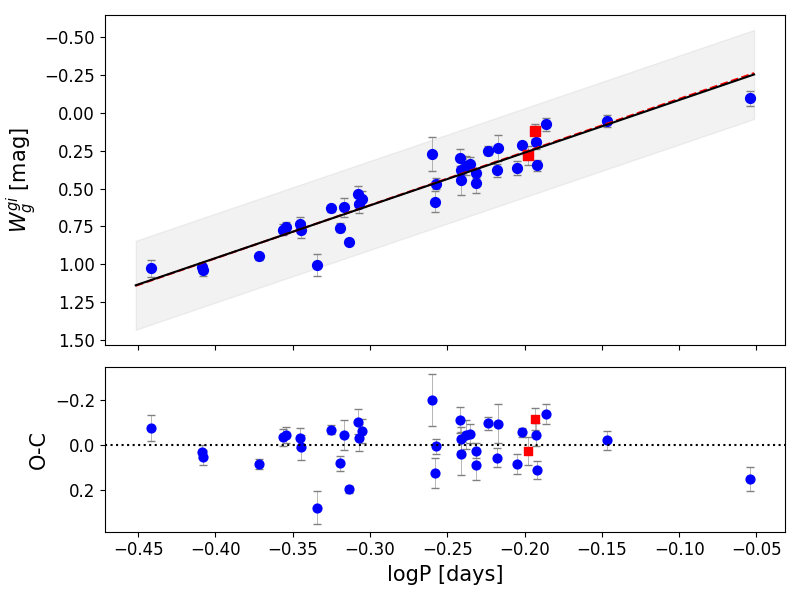}
    \caption{$W_{gi}$ RRL PLR. In red dashed line is the original regression from \citet{narloch24} in red dashed line, while the new PLR from Eq. \ref{eq:plr} is in solid black line. The two RRL in binary systems removed to form the new relation are shown in red symbols. The grey band indicates a 3-$\sigma$ confidence interval. The lower panel indicates the residuals against the new regression.}
    \label{fig:plr}
\end{figure}

Even though we have shown binarity is detected only in about 10\% to 15\% of RRL, with the majority of companions being significantly fainter than the RRL, it is worth reviewing what impact this may have in the period-luminosity relation (PLR) that these pulsators follow, since the addition of unaccounted light may systematically shift these relations like in the case of Cepheids \citep{karczmarek23}. The PLR of RRL is one of the main staples of the distance scale to population II stars \citep[e.g.][]{catelan04}, but the impact of binarity among RRL in their PLR remains unexplored.

In order to investigate this issue, we use the RRL sample of \citet{narloch24}, who constructed PLRs in the SDSS $g'r'i'$ filters based on 45 local RRL (38 RRab plus 7 RRc).  Twenty seven out of their 38 RRab stars are part of our speckle sample, and out of those, we identify  two RRL that host companions detected by speckle imaging, BH Peg and DN Aqr. Fig. \ref{fig:plr} shows the PLR based on the 38 RRab using  the reddening-free \citep{madore82} index $W_{gi}$, where PLR relations show the least scatter. The two RRL with companions are marked with red symbols. 

Removing these two stars, we re-calculate this PLR, obtaining

\begin{equation}\label{eq:plr}
    W_{gi}= (-3.477 \pm 0.208) (\log P - \log P_0) + (0.437 \pm 0.017),
\end{equation}
where $\log P_0=-0.25$ following \citet{zgirski23}. These new slope and intercept can be compared to those found originally by \citet{narloch24}, $a=-3.508\pm 0.202$ and $b=0.434\pm0.016$. Both relations are shown in Fig. \ref{fig:plr}. The new slope and intercept confirm that the impact of binarity in the PLR of RRL is negligible, and does not contribute to its intrinsic scatter, and therefore it is even less of an issue compared to Cepheids, which are know to have a much higher binary fractions \citep{kervella19a,evans20}, but also with almost negligible impact \citep{karczmarek23,narloch23}.  Larger samples of RRL with high quality parallaxes like those used by \citep{narloch24} would be needed to confirm this finding.

\section{Summary and conclusions}\label{sec:conclusions}

We conducted speckle imaging of 81 RRL in the solar neighborhood. Our observations reveal companions to 10 of them. Given the very small separations (at most 0.2\arcsec) and the very low probability of chance alignments we consider all of them as physical binaries, this therefore being the first direct imaging of RRL in binary systems. Very distant systems, with separations larger than about 500 AU, are found to be extremely scarce. Despite some ambiguous cases, the majority of detected companions are close to the upper MS or the low RGB of an assumed ancient ($\sim$10 Gyr) old population, and therefore have masses between 0.7 and 0.8 M$_{\odot}$. The binary fraction for RRL in the studied separation range is 12.3\%, ruling out a binarity fraction higher than 25\% based on binomial statistics. Metal-rich, thin disc RRL are found to have a lower binary fraction close to 6\%,  albeit a larger sample would be needed to confirm differences in the binary fraction as a function of metallicity.

Speckle interferometry significantly outperforms other methods in the search for RRL companions, but it comes with its own set of limitations, namely, that all detected companions are necessarily in wide orbits (>20 AU) and therefore the much sought after measurement of RRL masses would be only possible after many years of speckle follow-up. Secondly, that detections near the diffraction limit can have unreliable photometry (especially in bluer bandpasses like EO562) that muddles the interpretation of the nature of the companions.

We also studied the colors at minimum light for a selected sample of RRL, mostly as a validation of the speckle colors. RRL with bright enough companions, are visible in distinct locations in the color-color plane, making the minimum light colors a useful tool for the search and analysis of RRL in binary systems.

Finally, we revisited the PLR of RRL, in particular that in the reddening-free index $W_{gi}$, which shows the least scatter among the optical bands, and found that the impact of binaries,  albeit with a small sample, is negligible, changing neither the scatter nor the zero-point of the relation. 

\begin{acknowledgements}
In loving memory of my mother. We thank the referee for a thorough and speedy report. RSa acknowledges support from the European Research Council (ERC) under the European Union’s Horizon 2020 research and innovation program (grant agreement No. 951549 - UniverScale). Support for MC is provided by ANID's Proyecto Fondecyt Regular \#1231637 and ANID/Basal grant FB210003. RSm was supported by the National Science Center, Poland, Sonata BIS project 2018/30/E/ST9/00598". We also acknowledge support from the Polish Ministry of Science and Higher Education grant 2024/WK/02. Based on observations obtained at the international Gemini Observatory, a program of NOIRLab, which is managed by the Association of Universities for Research in Astronomy (AURA) under a cooperative agreement with the National Science Foundation. on behalf of the Gemini Observatory partnership: the National Science Foundation (United States), National Research Council (Canada), Agencia Nacional de Investigaci\'{o}n y Desarrollo (Chile), Ministerio de Ciencia, Tecnolog\'{i}a e Innovaci\'{o}n (Argentina), Minist\'{e}rio da Ci\^{e}ncia, Tecnologia, Inova\c{c}\~{o}es e Comunica\c{c}\~{o}es (Brazil), and Korea Astronomy and Space Science Institute (Republic of Korea). Observations in the paper made use of the High-Resolution Imaging instruments ‘Alopeke and Zorro. ‘Alopeke and Zorro were funded by the NASA Exoplanet Exploration Program and built at the NASA Ames Research Center by Steve B. Howell, Nic Scott, Elliott P. Horch, and Emmett Quigley.Observatory  Also based on data collected under the ESO/CAMK PAN – OCM agreement at the ESO Paranal Observatory. Software used in this work: astropy \citep{astropy13}, extinction \citep{barbary21}, LineOfSightBinaries \citep{neugent20}, matplotlib \citep{hunter07}, pyphot \citep{zenodopyphot}, scipy\citep{scipy}, SPISEA \citep{hosek20}.
\end{acknowledgements}

\bibliographystyle{aa}
\bibliography{rrl_speckle}

\begin{appendix}
\onecolumn

\begin{landscape}
\section{Target properties}\label{app:targets}
\scriptsize
\begin{longtable}{llcccrccccrrl}
\caption{Properties of the RRL sample and their observations}\label{tab:sample}\\
\hline
\hline
Name  & \textit{Gaia} DR3 ID & RA & Dec &  \textit{Gaia} G & Distance  & RUWE & Period & [Fe/H] & E(B-V) & JD & N sets & Instr\\
 & & J2000 & J2000 & [mag] & [pc] & & [days] & [dex] & [mag] & [days]&\\
\hline
\endfirsthead
\caption{continued.}\\
\hline\hline
Name  & \textit{Gaia} DR3 ID & RA(J2000.0) & Dec(J2000.0) &  G & Distance  & RUWE & Period & [Fe/H] & E(B-V) & JD & N sets & Instr\\
\hline
\endhead
\hline
\endfoot
AA Aql & 4224859720193721856& 20:38:15.07 & --02:53:25.50 & 12.04 & 1309.2 & 1.13 & 0.36179 & --0.58&0.05 & 2459510.80859 & 8 & a\\
AA CMi$^{*,+}$ & 3111925220109675136& 07:17:19.20 & +01:43:39.88 & 11.72 & 1093.1 & 1.07 & 0.47633 & --0.55&0.089& 2459506.12679 & 8 & a\\
AE Boo & 1234729400256865664 & 14:47:35.26 & +16:50:43.55 & 10.70 & 879.27 & 1.06 & 0.31489 & --1.62 & 0.023 & 2460840.75455 & 8 & a\\
AF Vel & 5360400630327427072& 10:53:02.50 & -49:54:23.00 & 11.50 & 1165.3 & 0.98 & 0.52741 & --1.64 &0.103& 2459214.80398 & 8 & z\\
AL CMi$^*$ & 3143813565573130880& 07:48:57.10 & +05:38:10.91 & 11.96 & 1570.4 & 2.50 & 0.55052 & --0.85 &0.035& 2459511.11708 & 8 & a\\
AM Vir & 3604450388616968576& 13:23:33.30 & --16:39:57.90 & 11.61 & 1368.3 & 1.25 & 0.61510 &--1.37 &0.047& 2459274.81271 & 8 & z\\
AR Per$^*$ & 234108363683247616 & 04:17:17.20 & +47:24:01.00 & 10.42 & 507.9 &1.07 & 0.42555 & --0.43 &0.206 & 2459187.86899 & 5 & a\\
AR Ser & 4427066402532783744 & 15:33:30.81 & +02:46:37.88 & 12.02 & 1700.0 & 1.16 & 0.57521 & --1.51 & 0.040 & 2460838.83361 & 7 & a\\
AT And & 1925406252226143104& 23:42:30.79 & +43:00:50.99 & 10.70 & 915.3 & 13.99 & 0.61691 & --0.97&0.111& 2459510.75125 & 8 & a\\
AT Vir & 3677686044939929728& 12:55:10.51 & --05:27:31.78 & 11.53 & 1250.8 & 1.28 & 0.52577& --1.91&0.025& 2459268.74609 & 8 & z\\
AV Peg & 1793460115244988800& 21:52:02.82 & +22:34:28.83 & 10.78 & 665.4& 1.25 & 0.39038 & --0.44 &0.062& 2459507.72435 & 10 & a\\
BB Eri & 2976126948438805760 & 04:53:37.55 & --19:26:00.90 & 11.40 & 1375.0 & 1.74 & 0.56991& --1.51& 0.023& 2459478.88209 & 8 & z\\
BB Vir & 3720468832650308352 & 13:51:40.78 & +06:25:51.39 & 11.20 &2754.3 & 4.26 & 0.47110 & --1.80 & 0.023 & 2460840.76829 & 11 & a\\
BH Aur$^*$ & 182142003881848832& 05:12:04.30 & +33:57:47.08 & 11.70 & 801.2 & 1.26 & 0.45608 & --0.34 &0.312& 2459504.08760 & 8 & a\\
BH Peg & 2828497068363486720& 22:53:00.97 & +15:47:15.59 & 10.53 & 852.3 & 1.22 & 0.64101 & --1.38&0.062& 2459510.76509 & 8 & a\\
BR Aqr & 2438710609949867776 & 23:38:32.91 & --09:19:07.01 & 11.49 & 1239.8 & 0.90 &0.48186 & --0.84&0.019& 2459416.79753 & 8 & z\\
\ditto & \ditto & \ditto & \ditto & 11.51 & \ditto & \ditto & \ditto & \ditto & \ditto & 2459480.71214 & 8 & z\\
CG Peg$^*$ & 1797739517580809856& 21:41:16.79 & +24:46:23.96 & 11.35 & 958.9 & 0.98 & 0.46714 & --0.48 &0.07& 2459507.73666 & 8 & a\\
CN Lyr$^{*,+}$ & 4539434124372063744 & 18:41:15.94 & +28:43:21.18 & 11.34 & 863.75 & 1.11 & 0.41138 & --0.08 & 0.175 & 2460836.96846 & 5 & a\\
CP Aqr & 2688734709383609344& 21:10:12.78 & --01:43:16.50 & 11.91 & 1298.7 & 1.04 & 0.46340 & --0.90&0.039 & 2459510.77796 & 8 & a\\
CZ Lac & 2000976545403561984 & 22:19:30.73 & +51:28:14.82 & 11.69 &1233.58 & 1.08 & 0.43212 & --0.50 & 0.241 & 2460839.08897 & 5 & a\\
DH Peg & 2720896455287475584 & 22:15:25.63 & +06:49:21.43 &  9.66 & 484.46 & 1.23 & 0.25551 & --1.35 & 0.082 & 2460839.02720 & 4 & a\\
DM And & 1912453760434108928& 23:32:00.71 & +35:11:49.31 & 12.01 & 1933.8 & 1.27 & 0.63042 & --2.32&0.07 & 2459510.81801 & 8 & a\\
DM Cyg$^*$ & 1853751148171392256& 21:21:11.48 & +32:11:29.00 & 11.63 & 1078.7 & 2.00 & 0.41986 & --0.14&0.111& 2459509.82961 & 8 & a\\
DN Aqr & 2381771781829913984& 23:19:17.19 & --24:12:58.95 & 11.23 &1278.1 & 0.95 & 0.63374& --1.63& 0.013& 2459416.87573 & 8 & z\\
DX Del$^+$ & 1760981190300823808& 20:47:28.39 & +12:27:51.05 & 10.05 & 568.2& 0.98 & 0.47261 & --0.56&0.067& 2459507.85133 & 8 & a\\
DZ Peg & 2817589255885467520 & 23:20:07.01 & +16:04:06.52 & 12.27& 1895.9 & 1.57 & 0.60737 & --1.52 & 0.056 & 2459501.87252 & 8 & a\\
FW Lup$^+$ & 6005656897473385600 & 15:22:25.38 & --40:55:36.23 & 9.01 & 356.4 & 1.08 & 0.48416 & --0.17 & 0.156 & 2460808.76819 & 4 & z\\
HH Pup$^{*,+}$ & 5510293236607430656 & 16:01:21.49 & +19:47:50.66 & 11.42 & 880.0 & 1.07 & 0.39074 & --0.69  & 0.081 &2459214.73479 & 8 & z\\
HK Pup & 3030561879348972544 & 07:44:46.81 & --13:05:54.63 & 11.42 & 1245.0 & 0.86 & 0.73419 & --1.26 & 0.091 & 2459215.64250 & 8 & z\\
IK Hya & 3486473757325180032& 12:04:47.18 & --27:40:44.28 & 10.17 & 774.1 & 1.23 & 0.65029& --1.36&0.033 & 2459268.73419 & 8 & z\\
IU Car & 5480600096166907776 & 06:53:07.52 & --59:35:43.42 & 11.56 & 1703.8 &1.17 & 0.73699&--1.85 &0.076 & 2459475.90696 & 8 & z\\
RR Cet & 2558296724402139392& 01:32:08.22 & +01:20:28.91 & 9.75 & 610.4 & 1.01 & 0.55302 & --1.52& 0.028 & 2459511.89637 & 8 & a\\
RR Gem & 886793515494085248& 07:21:33.50 & +30:52:59.09 &  11.68 & 1146.8& 1.57 & 0.39728 & --0.35&0.013 & 2459509.12189 & 12 & a\\
RU Cet & 2371321782802181888 & 01:00:40.32 & -15:57:28.37 & 11.84 & 1730.9 & 1.25& 0.58628& --1.60 & 0.015 & 2459480.81830 & 8 & z\\
RU Scl & 2336550174250087936 & 00:02:48.17 & --24:56:43.34 & 10.29 & 788.4 &  1.24 & 0.49336& --1.25 & 0.008 & 2459416.81559 & 8 &z\\
\ditto & \ditto & \ditto & \ditto & 10.27 & \ditto &  \ditto & \ditto & -\ditto & \ditto & 2459480.72655 & 8 & z\\
RV Cet & 5151789464548893184 & 02:15:14.93 & --10:48:01.28 & 10.91 & 1024.7 & 1.13 & 0.62342& --1.32 & 0.032 & 2459571.59442 & 8 & z\\
RV Phe & 6526559499016401408 & 23:28:31.57 & --47:27:13.37 & 12.00 & 1788.6& 1.44 & 0.59640&--1.60 & 0.010 & 2459480.78028 & 8 & z\\
RW TrA$^{*,+}$ & 5815008831122635520 & 17:00:46.68 & --66:39:50.32 & 11.46 & 918.9 & 1.00 & 0.37404 & +0.07 & 0.078 & 2459417.60510 & 8 & z\\
RX Cet & 2373827054405627904 & 00:33:38.31 & --15:29:14.88 & 11.47 & 1303.7 & 1.33 & 0.57369& --1.46 & 0.016 & 2459416.84368 & 8 & z\\
RX Eri & 2981136563934324224 & 04:49:44.28 & -15:44:28.10 & 9.69 & 579.9 & 1.28 & 0.58723&--1.33 & 0.031& 2459480.86456 & 8 & z\\
RZ Cet & 5176548537965653120 & 02:28:32.42 & --08:21:29.82 & 11.93 & 1579.8 & 1.24 & 0.51060&--1.50 & 0.02& 2459480.83008 & 8 & z\\
S Ara  & 5947570591534602240& 17:59:10.69 & -49:26:00.15 & 10.98 & 883.7 & 1.04 & 0.45188 & --1.43& 0.109& 2459415.64347 & 8 & z\\
SS For & 5117708899055276416 & 02:07:52.05 & --26:51:59.54 &10.43 & 775.7 & 1.27 & 0.49542& --1.35& 0.012& 2459474.74867 & 8 & z\\
SV Eri & 5165689383172441216 & 03:11:52.12 & --11:21:14.95 & 10.03 & 734.1 & 0.98 & 0.71386& --2.04& 0.037& 2459480.84117 & 8 & z\\
SV Hya & 3499611306368945536& 12:30:30.50 & --26:02:50.96 & 10.76 & 856.1 & 1.56 & 0.47855& --1.70 & 0.055& 2459268.75703 & 8 & z\\
SW And$^{*,+}$ & 2857456211775108480 & 00:23:43.09 & +29:24:03.66 & 9.81 & 501.6 &1.20 & 0.44225& --0.38 & 0.056&2459186.70671 & 10 & a\\
SW Aqr & 2689556491246048896& 21:15:17.82 & +00:04:33.63 & 11.31 & 1077.6 & 0.98 & 0.45930 & --1.24& 0.040& 2459511.79497 & 8 & a\\
SX Aqr & 2697816774212075904& 21:36:08.33 & +03:13:48.88 & 11.94 & 1620.2 & 1.76 & 0.53571 &--1.83 & 0.015 & 2459509.79300 & 8 & a\\
SZ Gem & 670266862434827776& 07:53:43.50 & +19:16:23.36 & 12.00 & 1426.5 & 1.44 & 0.50114 & --1.81& 0.037 & 2459502.11162 & 8 & a\\
SZ Hya & 5743059538967112576& 09:13:48.81 & --09:19:09.00 & 11.52 & 1306.0 & 1.12 & 0.53723 & --1.75&0.036& 2459214.78241 & 8 & z\\
TT Cnc & 603291108060958848& 08:32:55.11 & +13:11:28.19 & 11.46 & 1203.9 & 1.18 & 0.56345 & --1.58 & 0.034 & 2459190.09441 & 8 & a\\
\ditto & \ditto & \ditto & \ditto & 11.45& \ditto & \ditto& \ditto& \ditto & \ditto & 2459270.67076 & 8 & z\\
TT Lyn & 1009665142487836032 & 09:03:07.62 & +44:35:07.40 & 9.92 & 679.4 & 0.86 & 0.59742& --1.59 & 0.025&2459188.11998 & 8 & a\\
TU UMa & 4022618712476736896 & 11:29:48.48 & +30:04:02.37 &  9.94 & 626.12 & 1.00 & 0.55766 & --1.55 & 0.020 & 2460839.77884 & 5 & a\\
TV Lyn & 975895390891334656 & 07:33:31.72 & +47:48:10.02 & 11.54 & 1091.4 & 1.09 & 0.24185 & --0.99$\dagger$ &0.106 &2459188.10830 & 8 & a\\
TW Her$^{*,+}$ & 4596935593202765184 & 17:54:31.20 & +30:24:37.68 & 11.50 & 1117.3 & 1.04 & 0.39960 & --0.39 &0.037 & 2460836.93552 & 6 & a\\
TW Lyn & 924355886418231680 & 07:45:06.29 & +43:06:41.74 & 12.10 & 1487.8 & 1.25 & 0.48187& --1.23 & 0.036&2459188.09750 & 8 & a\\
TZ Aur$^*$ & 949205983077666944 & 07:11:35.01 & +40:46:37.29 &12.08 & 1349.0 & 1.71& 0.39167& --0.80 & 0.059&2459188.05873 & 8 & a\\
U Lep  & 2973463347160718976& 04:56:18.06 & --21:13:03.23 & 10.82& 1014.6 & 1.33 & 0.58147& --1.93&0.020& 2459480.87503 & 8 & z\\
U Pic$^+$  & 4784552718312266624 & 04:50:06.61 & --50:39:25.07 & 11.66 & 1216.4 &0.92 & 0.44037 & --0.73 & 0.013 & 2459480.79475 & 8 & z\\
UU Vir & 3698725337376560512 & 12:08:35.07 & --00:27:24.30 & 10.74 & 778.6 & 1.32 & 0.47562 & --0.82 & 0.016 & 2460807.65092 & 4 & z\\
UV Oct & 5768557209320424320 & 16:32:25.53 & --83:54:10.49 & 9.68 &  544.1 & 1.00 & 0.54257 & --1.61 & 0.086 & 2458685.55610 & 5 & z\\
\ditto & \ditto & \ditto & \ditto & 9.65 & \ditto & \ditto & \ditto & \ditto & \ditto & 2460900.50863 & 10 & z \\
UY Cyg$^*$ & 1858568795812429056 & 20:56:28.28 & +30:25:39.87 & 11.13 & 1038.8 & 0.95 & 0.56070 & --1.03 &0.149& 2459506.78779 & 8 & a\\
V0341 Aql & 4229920123678447232& 20:32:31.64 & +00:35:06.26 & 11.08 & 1125.4& 1.21 & 0.57802 & --1.37 &0.063& 2459503.75030 & 8 & a\\
V0413 CrA & 6730211038418525056& 18:47:57.61 & --37:44:23.05 & 10.68 & 841.4 & 1.12 & 0.58933 & --1.21 &0.087 & 2459415.61648 & 8 & z\\
V0440 Sgr & 6771307454464848768 & 19:32:20.80 & --23:51:12.75 &10.47 & 712.0& 1.11 & 0.47749& --1.47 & 0.092& 2459416.74350 & 8 & z\\
\ditto & \ditto & \ditto &\ditto  &10.54 &\ditto  &\ditto  &\ditto &\ditto &\ditto & 2459474.59500 & 8 & z\\
V0445 Oph$^*$ & 4352084489819078784& 16:24:41.21 & --06:32:29.79 & 10.97 &637.4 & 1.05 & 0.39703 & --0.23 & 0.238 & 2459271.87168 & 8 & z\\
V0674 Cen & 6120897123486850944& 14:03:24.10 & --36:24:20.0 & 11.46 & 1200.7 & 1.13 & 0.49399 & --1.53 & 0.047 & 2459268.86422 & 9 & z\\
V0675 Sgr & 4039386574037718528& 18:13:35.41 & --34:19:02.01 & 10.42 & 833.7 & 0.75 & 0.64231& --2.01 & 0.084 & 2459415.56352 & 8 & z\\
V0690 Sco & 4035521829393903744& 17:57:38.59 & --40:33:27.08 & 11.58 & 1168.0 & 0.81 & 0.49226 & --1.11 & 0.111 & 2459416.62830 & 8 & z\\
V1645 Sgr & 6680420204104678272& 20:20:44.48 & --41:07:06.06 & 11.51 & 1407 & 0.90 & 0.55297& --1.74 & 0.031& 2459415.78646 & 8 & z\\
V Ind & 6483680332235888896 & 21:11:29.75 & --45:04:29.95 & 10.23 & 661.7 & 1.00 & 0.47961& --1.50 & 0.018 &2459474.58057 & 8 & z\\
VW Scl & 4985455998336183168 & 01:18:15.03 & -39:12:45.87 & 11.30 & 1110.4 & 1.74 & 0.51091& --1.06 & 0.006& 2459480.80760 & 8 & z\\
W Tuc & 4709830423483623808 & 00:58:09.74 & --63:23:44.68 &11.62 & 1595.7 &0.98 & 0.64225& --1.64 & 0.011& 2459474.70149 & 8 & z\\
WY Ant & 5461994302138361728& 10:16:04.91 & --29:43:42.00 & 11.04 & 998.4 & 1.11 & 0.57436 & --1.66 & 0.042& 2459214.79328 & 8 & z\\
WZ Hya & 3765574712337027456& 10:13:24.10 & --13:08:17.57 & 10.96& 974.7 & 1.09 & 0.53772& --1.39 & 0.044& 2459268.76894 & 8 & z\\
X Ari  & 15489408711727488 & 03:08:30.88 & +10:26:45.16 & 9.62 & 536.1 & 1.22 & 0.65117& --2.40 & 0.254&2459188.84434 & 8 & a\\
X Crt & 3587566361077304704& 11:48:56.21 & --10:26:29.16 & 11.44 & 1398.3 & 1.33 & 0.73284 & --1.75 & 0.032& 2459268.80282 & 10 & z\\
XX And & 370067649378653440 & 01:17:27.39 & +38:57:02.17 & 10.73 & 1102.4 & 1.13 & 0.72276& --2.01 & 0.042 & 2459187.81050 & 5 & a\\
\ditto & \ditto & \ditto & \ditto & 10.81 & \ditto & \ditto & \ditto & \ditto & \ditto & 2459189.85728 & 8 & a\\
XX Pup & 5721192383002003200 & 08:08:28.24 & --16:31:59.52 & 11.55 & 1122.2 & 1.38& 0.51720& --1.42 & 0.046&2459228.67387 & 8 & z\\
XZ Cyg & 2142052889490819328 & 19:32:29.30 & +56:23:17.49 &  9.96 & 625.64 & 1.17 & 0.46660 & --1.50 & 0.109 & 2460839.08137 & 4 & a\\
Z Mic & 6787617919184986496 & 21:16:22.70 & --30:17:03.06 & 11.60 & 1217.3 & 0.96 & 0.58692 & --1.28 & 0.082 & 2460807.88767 & 6 & z\\
\end{longtable}
\tablefoot{ID, RA, Dec come from \textit{Gaia} DR3 \citep{gaiadr3}. Distances come from \citet{bailerjones21}. Periods and metallicities come from \citet{prudil20}, while reddenings come from \citet{lallement18}. Stars with a $^*$ sign in their names are part of the thin disc sample of \citet{prudil20}, while those with a \,$^{+}$ belong to the thin disc sample of \citet{abdollahi25} (see Sect. \ref{sec:thin}). $\dagger$ metallicity from \citet{kemper82}. JD indicates the Julian Date at the middle of the the set of observations. The last column indicates whether the observations were taken with 'Alopeke (a) or Zorro (z). Entries marked with the \ditto symbol are repeated observations of the same target on a different date.}
%\end{longtable}
\end{landscape}

\onecolumn
\section{Contrast curves and reconstructed images for the RRL sample}\label{app:contrast}

\begin{figure*}[h]
\centering
\begin{minipage}[]{0.45\textwidth}
\includegraphics[scale=0.45]{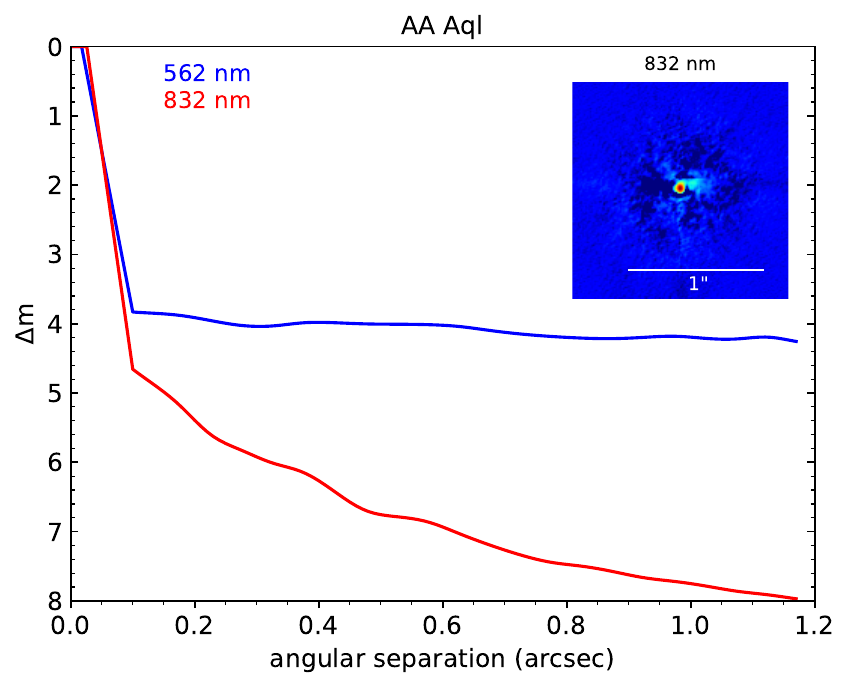}\\
\includegraphics[scale=0.45]{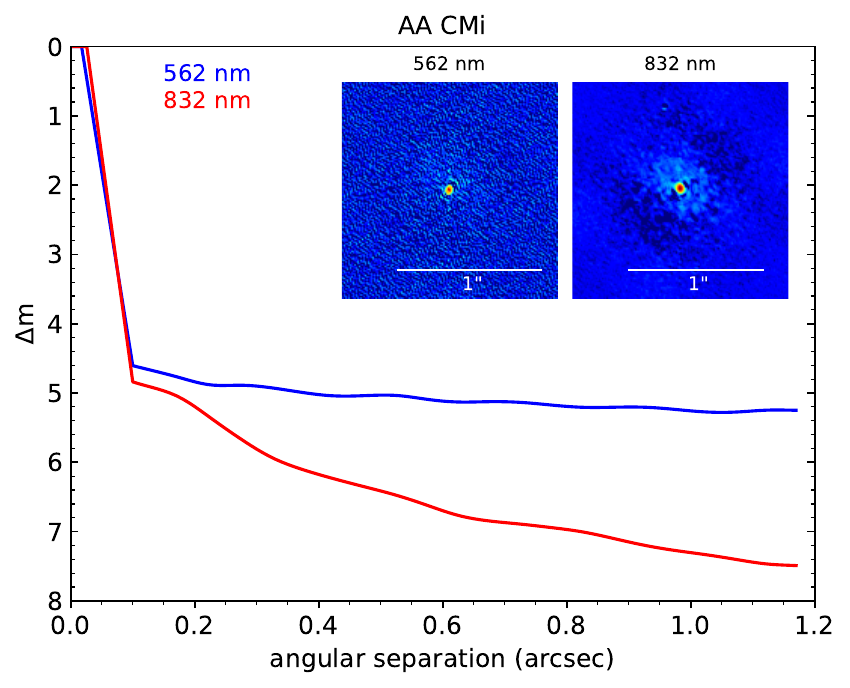}\\
\includegraphics[scale=0.45]{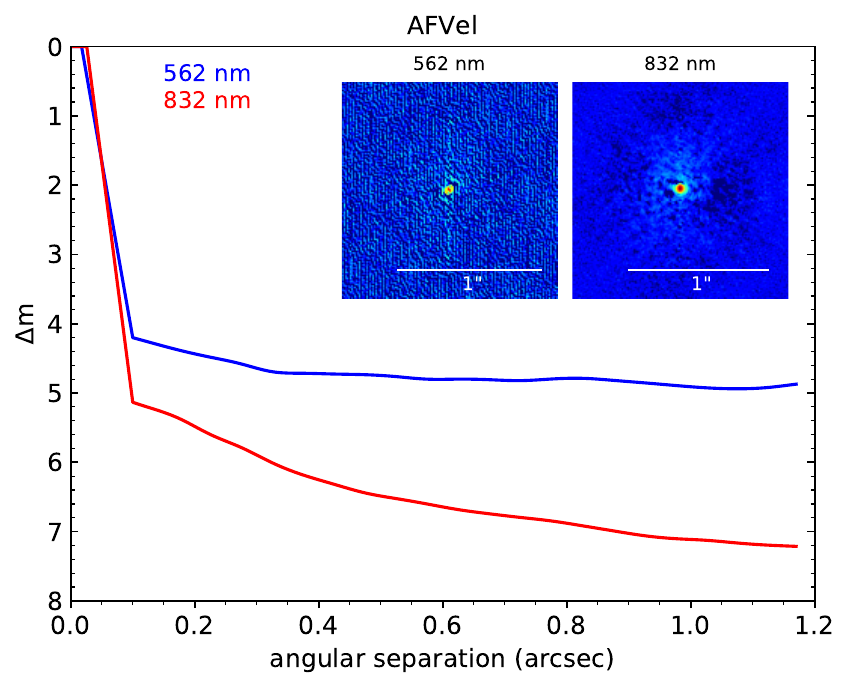}\\
\includegraphics[scale=0.45]{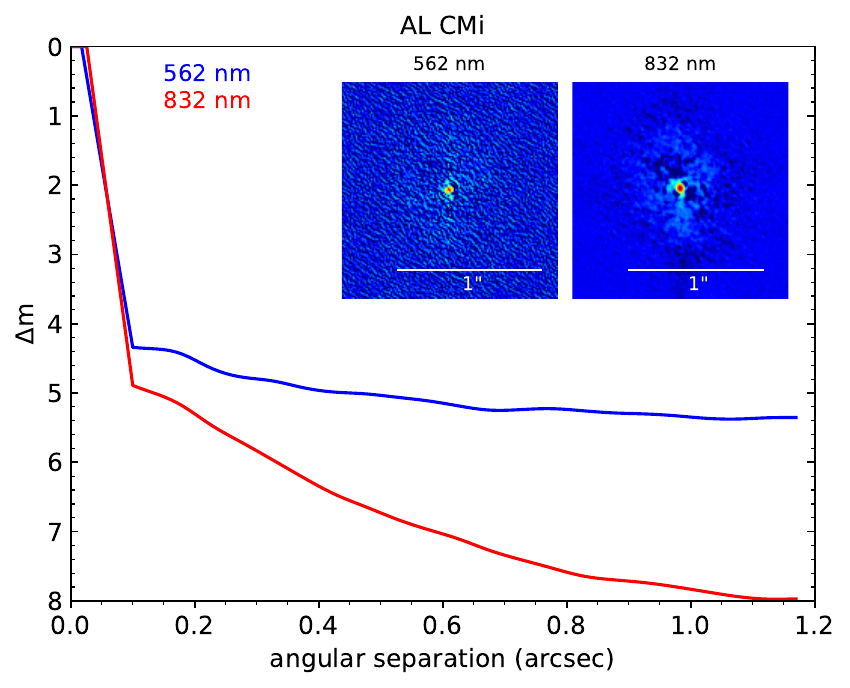}
\end{minipage}
\begin{minipage}[]{0.45\textwidth}
\includegraphics[scale=0.45]{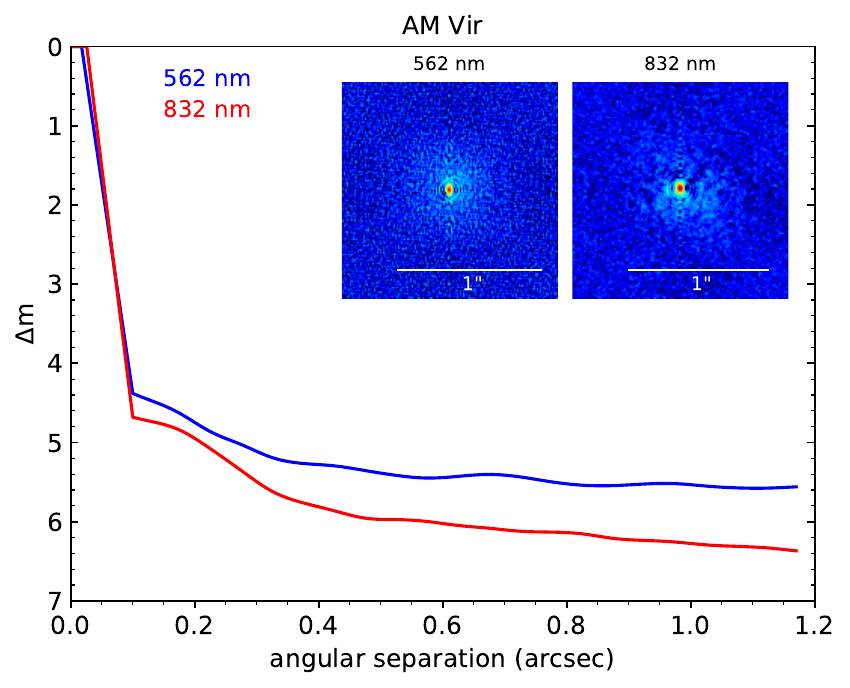}\\
\includegraphics[scale=0.45]{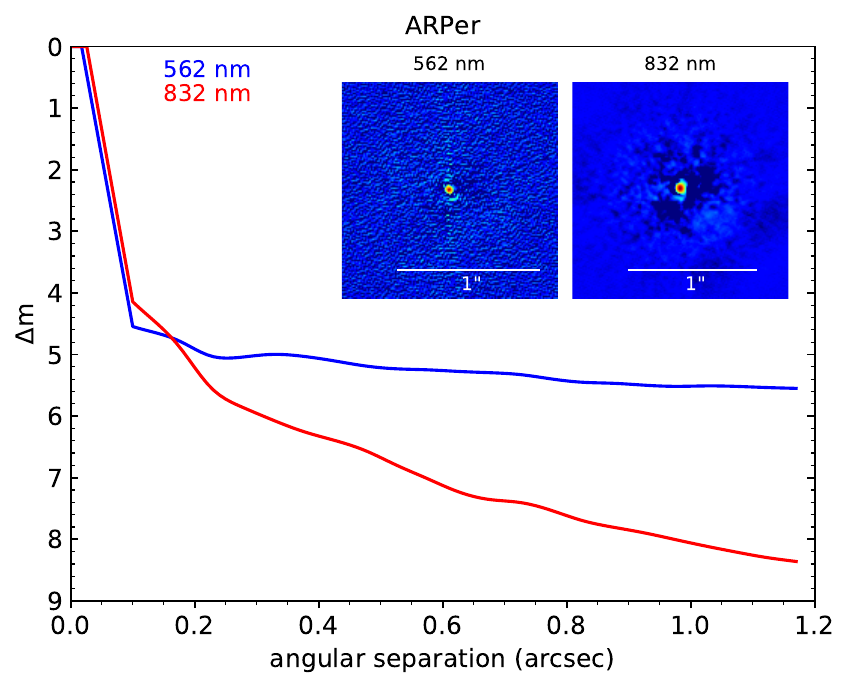}\\
\includegraphics[scale=0.45]{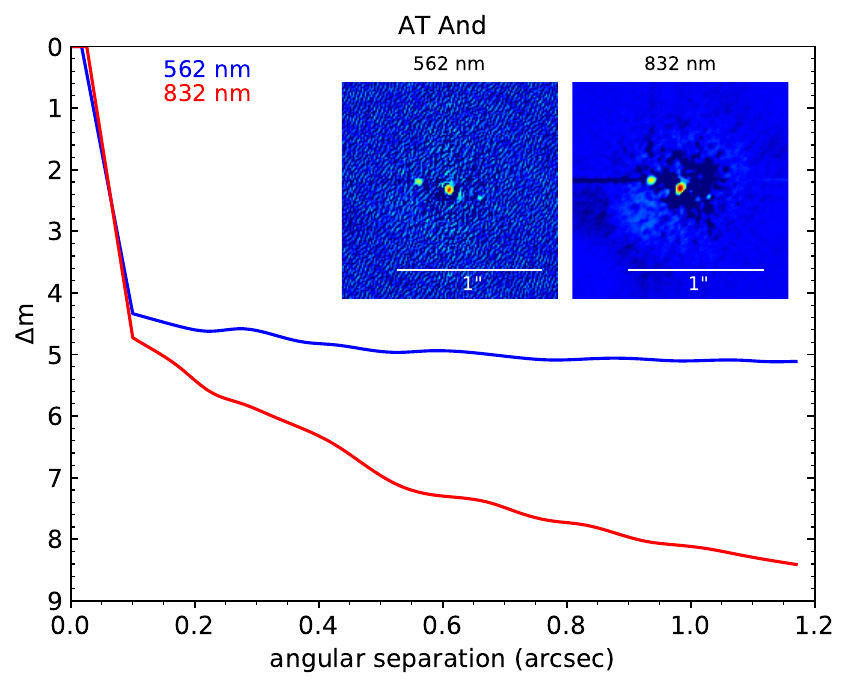}\\
\includegraphics[scale=0.45]{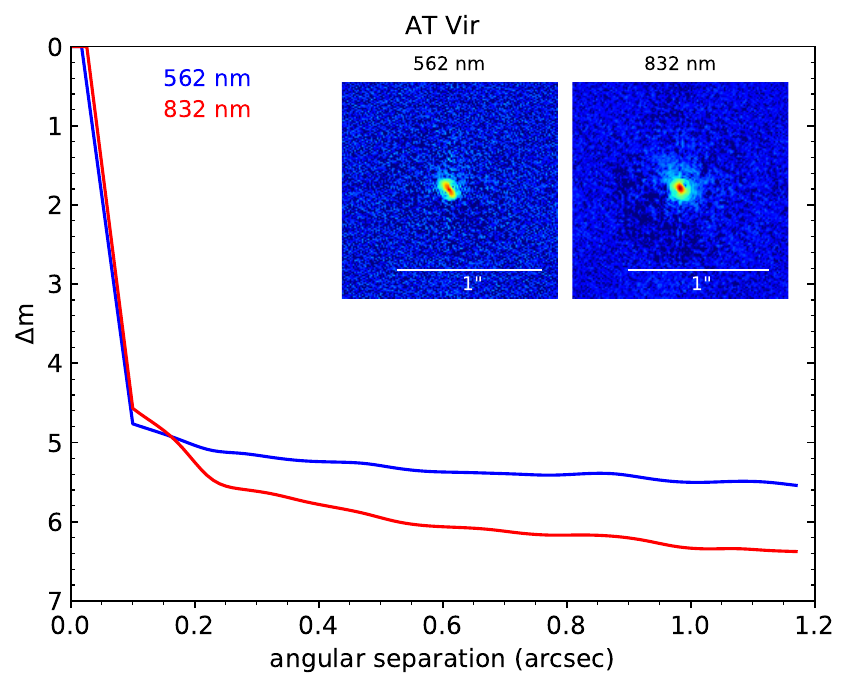}
\end{minipage} 
  \caption{Example 5-$\sigma$ contrast curves in the EO562 (blue lines) and EO832 filters (red lines), together with the reconstructed images in both filters shown as insets.}
    \label{fig:contrast1}
\end{figure*}

\onecolumn
\section{Speckle detection limits}\label{app:limits}

\scriptsize
\begin{longtable}{lcccccccc}
\caption{Summary of 5-$\sigma$ detection limits around the program stars.}\label{tab:contrast}\\
\hline
\hline
RR Lyrae  & \multicolumn{2}{c}{$\Delta m_{562}$} & \multicolumn{2}{c}{mass} & \multicolumn{2}{c}{$\Delta m_{832}$} &  \multicolumn{2}{c}{mass} \\
& \multicolumn{2}{c}{[mag]} &  \multicolumn{2}{c}{[M$_\odot$]} &  \multicolumn{2}{c}{[mag]} &  \multicolumn{2}{c}{[M$_\odot$]} \\
\cline{2-3} \cline{4-5} \cline{6-7} \cline{8-9}
  & 0.1\arcsec &  1.0\arcsec & 0.1\arcsec & 1.0\arcsec & 0.1\arcsec & 1.0\arcsec & 0.1\arcsec & 1.0\arcsec\\
  \hline
AA Aql & 3.83 & 4.19 & 0.79 & 0.76 & 4.65 & 7.75 & 0.68 & 0.30\\ 
AA CMi & 4.60 & 5.26 & 0.72 & 0.66 & 4.83 & 7.30 & 0.66 & 0.37\\ 
AE Boo & 5.16 & 6.39 & 0.64 & 0.54 & 5.20 & 8.57 & 0.61 & 0.21\\ 
AF Vel & 4.20 & 4.91 & 0.73 & 0.66 & 5.13 & 7.11 & 0.62 & 0.41\\ 
AL CMi & 4.34 & 5.35 & 0.74 & 0.64 & 4.89 & 7.83 & 0.66 & 0.33\\ 
AM Vir & 4.38 & 5.53 & 0.74 & 0.64 & 4.68 & 6.27 & 0.69 & 0.53\\ 
AR Per & 4.54 & 5.51 & 0.70 & 0.61 & 4.14 & 8.06 & 0.71 & 0.23\\ 
AT And & 4.34 & 5.10 & 0.77 & 0.70 & 4.72 & 8.11 & 0.71 & 0.33\\ 
AT Vir & 4.76 & 5.50 & 0.68 & 0.61 & 4.57 & 6.33 & 0.68 & 0.49\\ 
AV Peg & 5.45 & 7.36 & 0.63 & 0.48 & 5.00 & 8.55 & 0.62 & 0.19\\ 
BB Eri & 4.87 & 5.13 & 0.68 & 0.66 & 5.22 & 6.25 & 0.62 & 0.52\\ 
BB Vir & 4.89 & 6.49 & 0.81 & 0.70 & 4.89 & 8.17 & 0.81 & 0.51\\ 
BH Aur & 4.93 & 5.45 & 0.69 & 0.64 & 5.35 & 6.48 & 0.60 & 0.47\\ 
BH Peg & 5.41 & 7.47 & 0.65 & 0.48 & 4.70 & 8.25 & 0.70 & 0.30\\ 
BR Aqr & -- & -- & -- & -- & 4.92 & 6.45 & 0.65 & 0.50\\ 
\ditto & 4.44 & 5.25 & 0.73 & 0.65 & 5.04 & 7.79 & 0.64 & 0.32\\ 
CG Peg & 4.71 & 7.30 & 0.72 & 0.49 & 4.87 & 8.51 & 0.66 & 0.22\\ 
CN Lyr & 4.75 & 5.86 & 0.74 & 0.64 & 4.78 & 7.61 & 0.69 & 0.35\\ 
CP Aqr & 4.51 & 5.71 & 0.69 & 0.59 & 4.65 & 7.13 & 0.64 & 0.38\\ 
CZ Lac & 4.49 & 6.39 & 0.79 & 0.61 & 4.71 & 8.09 & 0.74 & 0.35\\ 
DH Peg & 4.43 & 6.63 & 0.70 & 0.52 & 4.46 & 9.10 & 0.68 & 0.16\\ 
DM And & 5.05 & 6.77 & 0.71 & 0.56 & 4.65 & 8.38 & 0.74 & 0.32\\ 
DM Cyg & 4.45 & 5.04 & 0.77 & 0.71 & 5.04 & 7.52 & 0.67 & 0.37\\
DN Aqr & --& -- & -- & -- & 4.39 & 7.14 & 0.72 & 0.43\\ 
DX Del & 4.02 & 4.86 & 0.79 & 0.71 & 4.73 & 8.26 & 0.70 & 0.26\\ 
DZ Peg & 4.97 & 5.81 & 0.70 & 0.62 & 4.95 & 7.02 & 0.69 & 0.47\\ 
FW Lup & 4.57 & 5.59 & 0.79 & 0.69 & 4.81 & 7.36 & 0.74 & 0.44\\ 
HH Pup & 5.01 & 6.41 & 0.66 & 0.55 & 4.81 & 6.07 & 0.64 & 0.50\\
HK Pup & 3.90 & 4.49 & 0.79 & 0.74 & 4.94 & 6.95 & 0.67 & 0.46\\ 
IK Hya & 4.66 & 6.85 & 0.71 & 0.53 & 4.81 & 7.47 & 0.68 & 0.41\\ 
IU Car & 3.95 & 4.02 & 0.76 & 0.76 & 4.65 & 6.23 & 0.69 & 0.53\\ 
RR Cet & 5.29 & 6.34 & 0.64 & 0.56 & 5.34 & 7.79 & 0.61 & 0.34\\ 
RR Gem & 5.10 & 6.60 & 0.71 & 0.58 & 4.47 & 7.85 & 0.74 & 0.31\\ 
RU Cet & 4.53 & 5.30 & 0.73 & 0.67 & 5.22 & 8.21 & 0.65 & 0.32\\ 
RU Scl & -- & -- & -- & -- & 4.50 & 7.38 &0.69 & 0.38\\ 
\ditto & 5.59 & 7.28 & 0.61 & 0.48 & 5.18 & 7.88 & 0.62 & 0.31\\ 
RV Cet & 4.50 & 5.09 & 0.72 & 0.66 & 5.25 & 6.73 & 0.62 & 0.47\\ 
RV Phe & 3.76 & 4.37 & 0.78 & 0.74 & 4.74 & 7.16 & 0.70 & 0.44\\ 
RW TrA & -- & -- & -- & -- & 4.70 & 6.59 & 0.73 & 0.49\\ 
RX Cet & -- & -- & -- & -- & 4.69 & 6.05 & 0.67 & 0.52\\ 
RX Eri & 5.13 & 7.33 & 0.64 & 0.47 & 5.35 & 8.79 & 0.59 & 0.19\\ 
RZ Cet & 4.39 & 5.49 & 0.72 & 0.62 & 5.03 & 7.26 & 0.64 & 0.41\\ 
S Ara & -- & -- & -- & --& 5.01 & 6.59 & 0.64 & 0.47\\ 
SS For & 5.00 & 5.82 & 0.66 & 0.58 & 5.11 & 6.71 & 0.62 & 0.45\\ 
SV Eri & 4.79 & 6.99 & 0.70 & 0.51 & 5.13 & 8.61 & 0.65 & 0.24\\ 
SV Hya & 4.82 & 6.38 & 0.67 & 0.54 & 5.05 & 7.46 & 0.62 & 0.36\\ 
SW And & 5.46 & 7.39 & 0.67 & 0.51 & 5.24 & 8.48 & 0.64 & 0.23\\ 
SW Aqr & 4.27 & 5.38 & 0.73 & 0.63 & 4.59 & 7.81 & 0.67 & 0.31\\ 
SX Aqr & 4.17 & 5.59 & 0.74 & 0.62 & 4.95 & 7.51 & 0.65 & 0.38\\ 
SZ Gem & 4.67 & 5.52 & 0.68 & 0.60 & 4.93 & 7.10 & 0.63 & 0.40\\ 
SZ Hya & 4.85 & 5.74 & 0.68 & 0.60 & 4.90 & 7.28 & 0.66 & 0.41\\ 
TT Cnc & 3.46 & 3.66 & 0.78 & 0.77 & 4.90 & 5.50 & 0.65 & 0.58\\ 
\ditto & 4.54 & 6.80 & 0.69 & 0.50 & 5.03 & 7.78 & 0.62 & 0.31\\ 
TT Lyn & 5.28 & 6.62 & 0.65 & 0.53 & 4.83 & 7.88 & 0.67 & 0.33\\ 
TU UMa & 4.40 & 7.53 & 0.71 & 0.46 & 4.34 & 8.89 & 0.71 & 0.19\\ 
TV Lyn & 4.18 & 4.80 & 0.79 & 0.73 & 4.87 & 7.18 & 0.69 & 0.43\\ 
TW Her & 5.07 & 6.52 & 0.71 & 0.59 & 4.93 & 7.87 & 0.69 & 0.32\\ 
TW Lyn & 4.81 & 5.34 & 0.67 & 0.63 & 4.74 & 7.00 & 0.65 & 0.41\\ 
TZ Aur & 4.79 & 5.94 & 0.67 & 0.57 & 4.98 & 7.35 & 0.61 & 0.34\\ 
U Lep & 5.32 & 7.53 & 0.63 & 0.45 & 5.06 & 7.90 & 0.63 & 0.30\\ 
U Pic & 4.32 & 5.28 & 0.72 & 0.64 & 5.19 & 7.54 & 0.60 & 0.34\\ 
UU Vir & 4.43 & 5.09 & 0.71 & 0.64 & 4.63 & 6.62 & 0.66 & 0.45\\ 
UV Oct & 3.49 & 4.01 & 0.78 & 0.75 & 4.15 & 4.84 & 0.73 & 0.66\\ 
'\ditto & 4.10 & 4.38 & 0.74 & 0.72 & 4.36 & 8.14 & 0.71 & 0.28\\ 
UY Cyg & 4.77 & 6.31 & 0.85 & 0.77 & 5.14 & 8.62 & 0.84 & 0.52\\ 
V0341 Aql & 4.42 & 5.23 & 0.74 & 0.66 & 4.61 & 7.25 & 0.71 & 0.44\\ 
V0413 CrA &-- & -- & -- &-- & 5.05 & 6.73 & 0.66 & 0.49\\ 
V0440 Sgr &-- & -- & --& -- & 5.05 & 7.94 & 0.62 & 0.29\\ 
\ditto & 5.30 & 7.31 & 0.62 & 0.46 & 5.30 & 7.31 & 0.59 & 0.38\\ 
V0445 Oph & 4.69 & 5.70 & 0.72 & 0.63 & 4.77 & 7.54 & 0.67 & 0.32\\ 
V0674 Cen & 5.09 & 5.92 & 0.66 & 0.58 & 5.33 & 6.97 & 0.60 & 0.44\\ 
V0675 Sgr & -- & -- & -- & -- & 5.13 & 7.17 & 0.65 & 0.43\\ 
V0690 Sco & --& -- & -- & -- & 5.43 & 6.42 & 0.60 & 0.50\\ 
V1645 Sgr & -- & --& -- & -- & 3.90 & 4.85 & 0.77 & 0.67\\ 
V Ind & 5.02 & 6.67 & 0.65 & 0.52 & 4.96 & 7.64 & 0.64 & 0.34\\ 
VW Scl & 4.68 & 5.88 & 0.69 & 0.58 & 5.06 & 7.54 & 0.63 & 0.35\\ 
W Tuc & 4.01 & 4.58 & 0.76 & 0.72 & 5.12 & 6.47 & 0.65 & 0.51\\ 
WY Ant & 5.09 & 6.12 & 0.66 & 0.57 & 5.21 & 7.72 & 0.62 & 0.35\\ 
WZ Hya & 4.32 & 4.91 & 0.72 & 0.67 & 5.22 & 7.17 & 0.62 & 0.41\\ 
X Ari & 5.11 & 6.58 & 0.69 & 0.56 & 4.97 & 7.49 & 0.69 & 0.43\\ 
X Crt & 5.12 & 6.01 & 0.67 & 0.59 & 5.49 & 7.33 & 0.60 & 0.41\\ 
XX And & 4.40 & 5.08 & 0.74 & 0.68 & 4.84 & 7.41 & 0.69 & 0.42\\ 
\ditto & 4.71 & 6.40 & 0.71 & 0.57 & 4.84 & 7.66 & 0.69 & 0.39\\ 
XX Pup & 4.63 & 5.59 & 0.69 & 0.60 & 4.90 & 7.36 & 0.64 & 0.38\\ 
XZ Cyg & 4.50 & 6.38 & 0.72 & 0.56 & 4.89 & 8.92 & 0.67 & 0.20\\ 
Z Mic & 4.38 & 4.86 & 0.72 & 0.67 & 5.05 & 6.52 & 0.63 & 0.48\\ 
\hline
\end{longtable}

\section{A reanalysis of proper common pairs}\label{app:cpm}
\begin{table*}[h]
\caption{Common proper motion (CPM) candidates from \citet{kervella19a} with updated DR3 analysis. Only RR Leo (\textit{Gaia} ID 630421935431871232, in bold below) remains as a CPM under our analysis using \textit{Gaia} DR3 astrometry.}           
\label{tab:cpm}      
\centering                          
\scriptsize
\begin{tabular}{lrrrrcccrrrccr}        
\hline\hline                
\multicolumn{6}{c}{{RRL}} & \multicolumn{5}{c}{{ Candidate companion}} & \multicolumn{3}{c}{{Pair properties}}\\
\cline{1-5} \cline{6-10} \cline{11-14}
 source\_id & G & $\varpi$ & $\mu_{\alpha}$ & $\mu_{\delta}$ & source\_id & G & $\varpi$ & $\mu_{\alpha}$ & $\mu_{\delta}$ & $\rho$ & a & $\Delta \varpi$\ & $\Delta \mu$\\
   & [mag] & [mas] & [mas\,yr$^{-1}$] & [mas\,yr$^{-1}$] &  & [mag] & [mas] & [mas\,yr$^{-1}$] & [mas\,yr$^{-1}$]  & [\arcsec] & [AU] &  &  \\
\hline
380489851379496320 & 11.3 & 0.89 & --5.03 & --7.82 & 380489851377596928 & 13.5 & 0.88 & --4.43 & --7.71 & 3.43 & 3871 & 22.2 & 22.2 \\
1805526190214362752 & 12.8 & 0.64 & --0.37 & 2.16 & 1805526185917504512 & 16.1 & 0.65 & --0.29 & 2.18 & 3.18 & 4976 & 1.5 & 1.5 \\
1385661079389302400 & 9.2 & 2.64 & --6.84 & 11.52 & 1385661079389302144 & 16.8 & 3.07 & --4.73 & 12.39 & 5.17 & 1958 & 3.4 & 3.4 \\
{\bf 630421935431871232} & 10.8 & 1.05 & --15.36 & --10.09 & 630421931138065280 & 18.0 & 1.06 & --15.43 & --10.04 & 9.64 & 9167 & 0.6 & 0.6 \\
6345324587928571648 & 11.7 & 0.87 & --1.94 & --32.51 & 6345324695303800192 & 17.3 & 0.78 & --0.68 & --33.09 & 2.67 & 3078 & 9.5 & 9.5 \\
6029835295710102400 & 13.5 & 0.66 & --2.68 & --1.75 & 6029835295648727168 & 19.4 & 1.30 & --4.56 & --4.60 & 12.21 & 18638 & 10.9 & 10.9 \\
4053550658972964608 & 11.2 & 1.07 & --9.87 & --12.23 & 4053551410541112192 & 15.9 & 0.87 & --10.53 & --12.52 & 29.59 & 27638 & 11.6 & 11.6 \\
\hline                                   
\end{tabular}
\end{table*}

\section{Examples light curves from ZB08 $griz$ photometry}
\begin{figure}[h]
\begin{minipage}{0.5\textwidth}
  \centering
  \includegraphics[width=.7\linewidth]{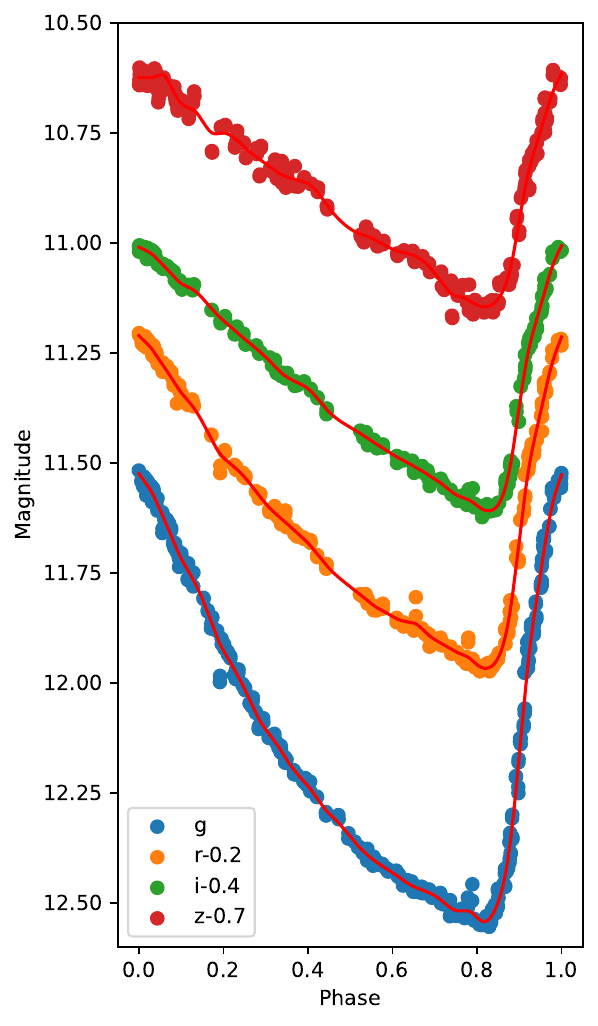}
\end{minipage}
\begin{minipage}{0.5\textwidth}
  \centering
  \includegraphics[width=.7\linewidth]{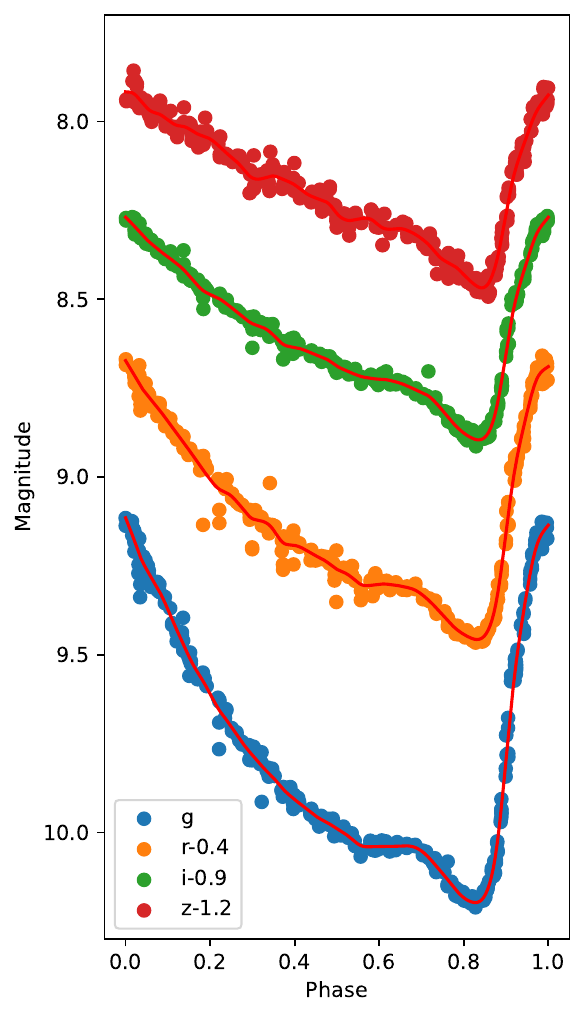}
\end{minipage}
\caption{$griz$ light curves for IU Car (left) and RX Eri (right), together with the spline interpolation used to determine the minimum light on each band (as a red solid line.)}\label{fig:lcs}
\end{figure}

\end{appendix}
\end{document}